\documentclass[12pt]{article}

\usepackage{multirow}
\usepackage{calc}
\usepackage{amsmath,amssymb,amsthm,amscd}
\numberwithin{equation}{section}
\usepackage[bf]{caption}
\usepackage{longtable}
\usepackage{array}
\usepackage{enumerate}
\usepackage{float}
\usepackage{subfig}
\usepackage{multicol}
\usepackage{multirow}
\usepackage{epsfig}
\usepackage{graphicx}
\usepackage{pict2e}
\usepackage{color}
\usepackage{authblk}
\usepackage{cite}
\usepackage[all]{xy}
\usepackage[width=1.09\textwidth]{caption}
\usepackage{bbm}
\usepackage{hyperref}

\usepackage[utf8]{inputenc}
\usepackage[T1]{fontenc}
\usepackage[english]{babel}
\usepackage{xspace}
\usepackage{pifont}

\newcommand{\U}[1]{\text{U(#1)}\xspace}
\newcommand{\SU}[1]{\text{SU(#1)}\xspace}
\newcommand{\SO}[1]{\text{SO(#1)}\xspace}

\def\cS{{\mathcal S}}

\newcommand{\one}[0]{\ensuremath{\mathbf{1} }\xspace}
\newcommand{\two}[0]{\ensuremath{\mathbf{2} }\xspace}

\newcommand{\three}[0]{\ensuremath{\mathbf{3} }\xspace}

\newcommand{\four}[0]{\ensuremath{\mathbf{4} }\xspace}
\newcommand{\six}[0]{\ensuremath{\mathbf{6} }\xspace}

\newcommand{\beq}{\begin{equation}}
\newcommand{\eeq}{\end{equation}}
\newcommand{\bea}{\begin{eqnarray}}
\newcommand{\eea}{\end{eqnarray}}

\newcommand{\nn}{\nonumber}

\begin{document}
\baselineskip=15pt
\begin{titlepage}
\begin{flushright}
\parbox[t]{1.73in}{CERN-PH-TH-2015-037}\\
\parbox[t]{.91in}{UPR-1271-T}\\
\parbox[t]{1.27in}{Bonn-TH-2015-02}
\end{flushright}

\begin{center}

\vspace*{ 0.0cm}
{\Large {\bf Three-Family Particle Physics Models from Global F-theory Compactifications}
}
\\[0pt]
\vspace{-0.1cm}
\bigskip
\bigskip {
{{\bf  Mirjam Cveti{\v c}}$^{\,\text{a}}$},
{{\bf  Denis Klevers}$^{\,\text{b}}$},
{{\bf  Dami\'an Kaloni Mayorga Pe\~na}$^{\,\text{c}}$},\\
{{\bf Paul-Konstantin~Oehlmann}$^{\,\text{d}}$},
{{\bf Jonas~Reuter}$^{\,\text{d}}$}
\bigskip }\\[3pt]
\vspace{0.cm}
{\it \small
${}^{\text{a}}$ Department of Physics and Astronomy,
University of Pennsylvania,\\ Philadelphia, PA 19104-6396, USA\\[8pt]
${}^{\text{b}}$Theory Group, Physics Department, CERN, CH-1211, Geneva 23, Switzerland\\[8pt]
${}^{\text{c}}$ The Abdus Salam International Centre for Theoretical Physics,
Strada Costiera 11, 34151, Trieste, Italy\\[8pt]
${}^{\text{d}}$ Bethe~Center~for~Theoretical~Physics, Physikalisches~Institut~der~Universit\"at~Bonn,\\ Nussallee~12,~53115~Bonn,~Germany\\[0.5cm]}

{\small cvetic@hep.upenn.edu, Denis.Klevers@cern.ch, dmayorga@ictp.it,
oehlmann@th.physik.uni-bonn.de, jreuter@th.physik.uni-bonn.de}
\\[0.4cm]
\end{center}

\begin{center} {\bf Abstract } \end{center}
\vspace{0.4Em}

We construct four-dimensional, globally consistent F-theory models with three chiral generations, whose
gauge group and matter representations coincide with those of the Minimal Supersymmetric Standard
Model, the Pati-Salam Model and the Trinification Model. These models result from compactification on
toric hypersurface fibrations $X$ with the choice of base $\mathbb{P}^3$.  We observe
that the F-theory conditions on the $G_4$-flux restrict the number of
families to be at least three. We comment on the phenomenology of the models, and for Pati-Salam and
Trinification models discuss the Higgsing to the Standard Model.
A central point of this work is the construction of globally consistent $G_4$-flux. For this purpose we
compute the vertical cohomology $H_V^{(2,2)}(X)$ in each case and solve the conditions imposed 
by matching the M- and F-theoretical 3D Chern-Simons terms.  We explicitly check that the expressions found
for the $G_4$-flux allow for a cancelation of D3-brane tadpoles. We also use the integrality of 3D Chern-Simons terms to ensure that our $G_4$-flux solutions are adequately quantized.

\begin{flushright}
\parbox[t]{1.2in}{March, 2015}
\end{flushright}
\end{titlepage}
\clearpage
\setcounter{footnote}{0}
\setcounter{tocdepth}{2}
\tableofcontents
\clearpage

\section{Introduction}
The construction of fully fledged particle physics models which reproduce the phenomenology of the Standard Model, while 
providing generic predictions for its behavior at higher scales remains a very active part of the research in string theory. One 
active front in this direction is F-theory \cite{Vafa:1996xn}, where the geometrization of certain properties of non-perturbative 
Type IIB string theory allows for a very systematic understanding and engineering 
of the gauge symmetry, particle content and  the interactions in a given model.

Most model building endeavors in F-theory have appealed to an underlying \SU5 gauge group 
(for a non exhaustive list of works see 
\cite{Donagi:2008ca,Beasley:2008dc,Beasley:2008kw,Donagi:2008kj,Marsano:2009ym,Blumenhagen:2009yv,Grimm:2009yu,Dudas:2009hu,Marsano:2009wr,Dudas:2010zb,Dolan:2011iu,Mayrhofer:2012zy,Braun:2013yti,Cvetic:2013uta,Krippendorf:2014xba}). 
This is partly due to the simple group theoretical embedding of the standard model gauge group and its representations into 
\SU5, with its well earned merit for gauge coupling unification\footnote{For a detailed discussion on gauge coupling unification 
in \SU{5} GUTs see e.g. \cite{Blumenhagen:2008aw,Dolan:2011aq}}. In addition this gives the advantage of having the full 
gauge symmetry concentrated on a single divisor in the base that allows for a local treatment of certain features of the model 
\cite{Donagi:2008ca,Beasley:2008dc,Beasley:2008kw,Donagi:2008kj}. Nevertheless, the increasing understanding of many 
global issues has prompted interest in   
alternative models which aim either at the direct construction of the bare MSSM\footnote{see \cite{Choi:2010su,Choi:2013hua} 
for earlier attempts to get the standard model gauge group from a deformation of the \SU5 singularity.} 
\cite{Lin:2014qga,Klevers:2014bqa,Grassi:2014zxa}, as well as alternative grand unification schemes such as the Pati-Salam 
model or 
Trinification, among others. These models have the advantage that they do not suffer from the pathological group theoretical 
puzzles inherent to \SU5, such as the doublet triplet splitting problem. These schemes constitute also a very promising 
alternative in other corners of the string landscape such as perturbative Type IIA/B and the heterotic string (see e.g. 
\cite{Cvetic:2004ui,Forste:2004ie,Kobayashi:2004ya}).

In the model building program one aims at reproducing certain features of the particle physics models such as appropriate 
gauge symmetries, particle representations, the right number of generations and, at least, the possibility to generate a 
hierarchy in the Yukawa textures (for reviews on this topic in an F-theory context see e.g. 
\cite{Weigand:2010wm,Maharana:2012tu}). As for the gauge symmetries, the appearance of non-Abelian factors has been 
understood since the beginning of F-theory and can be tracked by the degeneration of the elliptic fiber over codimension one 
surfaces on the base \cite{Morrison:1996na,Morrison:1996pp,Bershadsky:1996nh}, see \cite{Katz:2011qp} for recent 
refinements. Abelian gauge symmetries are due to the 
presence of sections of the elliptic fibration, in addition to the so-called zero section of the Weierstrass model 
\cite{Morrison:1996pp,Aspinwall:1998xj,Aspinwall:2000kf}. Being related to global objects, as well as being essential tools for 
controlling the phenomenology of particle physics models, \U1 symmetries have pushed the F-theory model building 
program towards a more global picture.  Similarly, the charged matter representations can be tracked by degenerations of the 
elliptic curve at codimension two in the base \cite{Katz:1996xe}, see also \cite{Morrison:2011mb} for the discussion
of higher symmetric representation. In order to achieve a chiral 
spectrum, the addition of $G_4$-flux is necessary,
see  \cite{Marsano:2011hv,Krause:2012yh} for first global examples.
The intersection of matter curves at codimension three leads to the geometrically allowed couplings of the model. In F-theory 
the hierarchy for the Yukawas is possible since generically, the Yukawas for one family are generated geometrically while for 
the other two of these couplings arise from instanton or flux contributions that are significantly small.

The compactification spaces which are commonly used for constructing 4D $\mathcal{N}=1$ effective theories in
F-theory are genus one fibered Calabi-Yau (CY) fourfolds. The global model building process is divided into two steps, the first
step being the construction of appropriate compactification manifolds exhibiting the desired fiber degenerations which lead to 
the appropriate gauge symmetry, matter and interactions. The second step is concerned with the construction of appropriate
$G_4$-flux to account for the desired chirality in the spectrum.

Regarding the construction of suitable CY-fourfolds, the efforts have been divided in two fronts: The first pushes towards the 
classfification and construction of all admissible bases \cite{Morrison:2012np,Morrison:2012js,Martini:2014iza}, see also 
\cite{Heckman:2013pva,DelZotto:2014hpa,DelZotto:2014fia,Heckman:2015bfa} for the related study of non-compact bases. 
The second is 
more concerned with the construction of genus-one or elliptic fibers $\mathcal{C}$ which naturally allow for certain generic 
features of the compactification. In this direction, there have been two major conceptual extensions: One is related to the 
construction of elliptic curves exhibiting an ever growing number of rational points 
\cite{Aldazabal:1996du,Klemm:1996hh,Klemm:2004km,Grimm:2010ez,Krause:2011xj,Grimm:2011fx,Park:2011ji,Esole:2011cn,Cvetic:2012xn,Mayrhofer:2012zy,Braun:2013yti,Borchmann:2013jwa,Cvetic:2013nia,Grimm:2013oga,Braun:2013nqa,Cvetic:2013uta,Borchmann:2013hta,Cvetic:2013jta,Cvetic:2013qsa,Mayrhofer:2014opa}. 
These permit the construction of elliptically fibered CY-manifolds with a certain number of rational sections and hence, a non 
trivial  Mordell-Weil (MW) group of rational sections. While the free part of the MW-group yields the U(1)-gauge fields in
F-theory \cite{Morrison:1996na}, the torsion part is responsible for the presence of non-simply
connected gauge groups \cite{Aspinwall:1998xj} and its effects are seen at codimension two as it forbids certain 
representations to be part of the theory. The other conceptual extensions has to do with fibers which entirely lack rational 
points. These lead to genus-one fibrations which do not have any section 
\cite{Braun:2014oya,Morrison:2014era,Anderson:2014yva}. Nevertheless, this type of fibrations is suitable for F-theory as their 
associated Jacobian fibration does have a section and describes the same physics. In the genus-one fibrations, the presence 
of an $m$-sections has been unveiled as the geometric object
responsible for the presence of discrete gauge symmetries in their associated effective SUGRA theory \cite{Klevers:2014bqa,Mayrhofer:2014haa,Garcia-Etxebarria:2014qua,Mayrhofer:2014laa,Cvetic:2015moa}.

There is a natural framework which provides the simplest examples for the two kinds of fibers described above: the 16 
inequivalent 2D toric varieties \cite{Kreuzer:1995cd,Grassi:2012qw}. By describing the genus-one fiber as an algebraic curve 
in any of these toric ambient spaces one ends up with any of the possible cases of elliptic fibers with up to four independent rational points as well as genus-one curves with two- and three-sections. In \cite{Klevers:2014bqa}, a systematic analysis of 
the effective six dimensional theories stemming from compactification of F-theory
on any of these 16 toric hypersurface fibrations over an arbitrary complex-two dimensional base was done. Abandoning the 
paradigm of the holomorphic zero section and allowing it to be simply rational, all resolution divisors 
inherited from the corresponding ambient space descend to Cartan divisors on the corresponding CY-manifold. 
In this fashion, it was possible to deduce the intrinsic gauge symmetry of each toric hypersurface fibration, without the 
necessity to introduce further specializations of the geometry such as tops \cite{Candelas:1996su,
Candelas:1997eh,Bouchard:2003bu}, which appear in algorithmic approaches to F-theory model building
\cite{Braun:2013nqa}. Among these intrinsic gauge symmetries, groups which are familiar for particle physicists such as 
$\SU3\times\SU2\times\U1$, $\SU4\times\SU2^2/\mathbb{Z}_2$ and $\SU3^3/\mathbb{Z}_3$ appear naturally. Even more 
interestingly, the matter representations that are possible in each of these models coincide with those needed for the MSSM, 
Pati-Salam model and Trinification, respectively. While the analysis of \cite{Klevers:2014bqa} was made in six dimensions, 
these findings carry over to 4D as well. Another important feature of the six-dimensional models is the existence of a so 
called "toric Higgsing" (or a chain of those) which allows for a pictorial, toric interpretation of Higgsings from
the Pati-Salam and Trinification models down to the MSSM.

In this work we study the four dimensional-version of the MSSM, the Pati-Salam model and Trinification engineered by 
F-theory on toric hypersurface fibrations $X$. For the first time we construct explicit, globally consistent, three-family models 
with the chiral matter content of the MSSM, Pati-Salam model and Trinification. These models result from three different toric 
hypersurface fibrations over a fixed base $B=\mathbb{P}^3$. We construct the vertical cohomology $H^{(2,2)}_V(X)$ and 
provide the most general expression for the $G_4$-flux in each case. These expressions are shown to comply with all 
conditions on M-theory Chern-Simons terms imposed by duality with F-theory and allow for a D3-brane tadpole canceling 
solution which involves an integral and positive number of D3-branes. Since the determination of an integral basis for the 
vertical cohomology cohomology $H^{(2,2)}_V(X)$, $G_4$-flux quantization is checked indirectly by ensuring quantization of 
the induced 3D Chern-Simons terms. Regarding the phenomenology of the considered models we have two main drawbacks: 
Since we have no control over the vector-like sector of the theory, it is not guaranteed that we have the vector-like pairs 
needed for the breaking of  electroweak symmetry neither for the breaking of the Pati-Salam or Trinification groups down to the 
Standard Model gauge group. In addition, we have only stated the trilinear couplings which are generated geometrically and 
have argued that under the assumption of the presence of a light pair of Higgses, these would allow for the generation of a 
hierarchy. However, the currently applicable tools do not allow us to provide to perform a quantitative analysis.

This paper is organized as follows: In Section \ref{sec:review} we summarize the general procedure to construct 4D chiral 
models from F-theory and shortly review the tools needed for our analysis: the general features of elliptic fibrations, $G_4$-flux 
and $G_4$-flux consistency conditions in F-theory. In Section \ref{sec:standardmodel} we discuss the elliptic fibration leading 
to the gauge group and matter content of the MSSM. There, we also compute $G_4$-flux and the resulting 4D matter 
chiralities. Simila as in the standard model, we see that the exact cancelation of anomalies (without Green Schwarz 
counterterms for anomalies involving $\U1_Y$) enforces a family structure. In this regime we scan over all allowed strata in the 
moduli space of the CY-manifold $X_{F_{11}}$ with base $\mathbb{P}^3$ and compute the smallest number of families for 
which the D3 brane tadpole is canceled with a positive integral number of D3-branes. We observe that three is in fact the 
smallest permitted number of families in this model. The same observation holds also for Pati-Salam and Trinification models. 
In addition we give closed formulas for the Hodge structure of all fibrations for the choice of base  $B=\mathbb{P}^3$ in 
appendix \ref{app:HodgeNumbers}. We generically find $h^{(2,1)}=0$  which restricts cosmological applications.
We conclude this section with a brief discussion of the phenomenology of the model under the assumption that a light pair of 
Higgs fields is present. In Sections \ref{sec:pati} and \ref{sec:trinif} we present a similar discussion of the Pati-Salam and 
Trinification models  as in Section \ref{sec:standardmodel}. In addition, we comment on the Higgsing down to the 
Standard Model gauge group. We indeed find that there exist three-family models both for Pati-Salam and Trinification,
that Higgs down to three-family Standard models.
Finally in section \ref{sec:conclusion} we present our conclusions and discuss future possible directions of research.
Appendix \ref{app:HodgeNumbers} contains a brief account on the computation of Hodge numbers of CY-fourfolds given
as toric hypersurfaces and applications to the considered cases $X_{F_{11}}$, $X_{F_{13}}$ and $X_{F_{16}}$ with base 
$B=\mathbb{P}^3$. Appendix \ref{app:polytope} contains the explicit lattice polytopes for the five-dimensional toric
ambient spaces of all considered toric hypersurface fibrations.

\section{Tools \& Strategies for Four-Dimensional Model Building}\label{sec:review}
In this section, which serves as a preparation for the analyses in Sections 
\ref{sec:standardmodel}, \ref{sec:pati} and \ref{sec:trinif}, we outline the basic  techniques for building  F-theoretic models of particle 
physics. 
Although many of the presented methods
are applicable to general Calabi-Yau (CY) fourfolds $X$, we focus here on the case of the toric 
hypersurface fibrations as studied in \cite{Klevers:2014bqa}.  Except for the discussion of $G_4$-flux 
quantization, this section is mainly a concise review on the construction
of chiral 4D F-theory models, following closely \cite{Cvetic:2013uta} to which
we refer for further details. 
In an accompanying Appendix \ref{app:HodgeNumbers} we discuss
the computation of Hodge numbers of  CY-fourfolds $X$ given as toric hypersurfaces.
The reader interested in the phenomenological results can safely skip this
section and continue with Section \ref{sec:standardmodel}.

\paragraph{F-theory Geometry:} On the geometry side, the starting point for the construction of an
F-theory model is the choice of a three-dimensional base manifold $B$ as well as the genus-one or 
elliptic fiber of the Calabi-Yau fourfold $X$. As a next step  all codimension one, two and three 
singularities of
the fibration have to be analyzed in order to determine the gauge group $G$, matter content, 
i.e.~the representations and their matter curves in $X$, and the Yukawa couplings of the 4D
effective theory of F-theory.  

In order to being able to construct $G_4$-flux we have to compute the
cohomology ring of $X$. For a CY-fourfold given as a toric hypersurface 
(or complete intersection) fibration, which is the case of interest of this work, 
the full cohomology ring of $X$ as a quotient polynomial ring generated by $H^{1,1}(X)$
\cite{Mayr:1996sh,Klemm:1996ts,Grimm:2009ef,Cvetic:2013uta}.
Concretely, for a CY-fourfold $X$ with a given toric base $B$, we choose a basis $D_A$, 
$A=0,\ldots,h^{(1,1)}(X)-1$, for the divisor group $H^{1,1}(X)$. Then the cohomology ring of $X$ is 
given by the polynomial ring in the $D_A$ divided by the Stanley-Reisner (SR) ideal of the ambient toric 
variety of $X$. 

As our focus is on chirality inducing $G_4$-flux, we are 
primarily interested in the subgroup $H^{2,2}_{\mathrm{V}}(X)$ of the fourth 
cohomology of $X$, the primary vertical cohomology 
\cite{Greene:1993vm}. It is given by the quotient ring at grade two, that is constructed by forming all 
possible products $D_A\cdot D_B$. These are linearly dependent. Thus, we compute the rank of the 
inner product matrix on these elements, which yields the dimension $h^{(2,2)}_V(X)$, and choose an 
appropriate basis. The topological metric on this basis is denoted by $\eta^{(2)}$.

We emphasize that the full Chern class of the CY-manifold $X$ can be computed independently 
of the base $B$ by using the adjunction formula and the total Chern class of the ambient space, see 
\cite{Cvetic:2013uta} for more details.
Of particular interest for F-theory are the second Chern class $c_2(X)$ and the Euler number
$\chi (X)$ of $X$.

\paragraph{Constraints on $G_4$-flux in F-theory:} 

F-theory on $X\times S^1$ and M-theory on $X$ are dual to each other \cite{Witten:1996bn}.
Thus, consistent $G_4$-flux in a four-dimensional F-theory compactification on $X$ is 
understood as $G_4$-flux in the dual M-theory compactification on the same $X$ to three dimensions, 
so that the $G_4$-flux  obeys additional consistency conditions. These consistency conditions follow 
from requiring that the three-dimensional 
effective actions of F- and M-theory agree, which can be used to derive the full effective action of 
F-theory \cite{Grimm:2010ks}.

In M-theory, $G_4$-flux has to fulfill two basic conditions. First, it must obey the following quantization condition \cite{Witten:1996md}:
\begin{align}\label{eq:quantization}
G_4+\frac{c_2 (X)}{2} \in H^{4}(X,\mathbb{Z}) \; .
\end{align}
Second,  the cancelation of M2-brane tadpoles, which lift to D3-brane tadpoles in Type IIB strings and F-theory, requires the equality \cite{Sethi:1996es, Gukov:1999ya}
\begin{align}
\frac{\chi (X)}{24}=n_{\rm D3}+\frac{1}{2}\int_X G_4\wedge G_4 \; ,
\label{eq:tadpole}
\end{align}
where  $n_{\rm D3}$ denotes the number of D3-branes. As mentioned before, we will focus here on
special $G_4$-flux that is entirely in the subgroup $H^{(2,2)}_V(X)$.\footnote{For
recent analyses of horizontal $G_4$-flux in F-theory, see 
\cite{Grimm:2009ef,Jockers:2009ti,Intriligator:2012ue,Bizet:2014uua}.}

For compatibility with the duality between M- and F-theory,  we need to impose additional 
conditions on the $G_4$-flux. These are most easily formulated in terms of conditions on
the Chern-Simons (CS) terms for the three-dimensional vectors on the Coulomb branch of the effective 
action  of the M-theory compactification on the CY-fourfold $X$. On the M-theory side, these CS-terms  are given by \cite{Haack:2001jz}
\begin{align}\label{eq:thetaM}
\Theta^\mathrm{M}_{AB}&=\int_{X} G_4\wedge D_A\wedge D_B\;,
\end{align}
where here and in the following, Poincar\'e duality is always understood. We note that the 3D CS-terms
have obey the quantization condition $\Theta^M_{AB}\in \mathbb{Z}$ or $\mathbb{Z}/2$, see 
e.g.~\cite{Belov:2005ze,Kapustin:2010hk} for recent discussions. We note that
these quantization conditions are expected to be equivalent to the  $G_4$-flux  quantization conditions
\eqref{eq:quantization}  \cite{Intriligator:2012ue}.

In the dual F-theory side the same CS-terms, denoted now by $\Theta^\mathrm{F}_{AB}$, have two 
contributions.  First, we can have classical CS-terms $\Theta^F_{\text{cl},\, AB}$, which either 
descend from 4D to 3D from gaugings of axions  or which correspond to circle fluxes 
\cite{Grimm:2011sk}. Second, CS-terms on the 3D Coulomb branch receive one-loop corrections 
from integrating out massive fermions \cite{Niemi:1983rq,Redlich:1983dv,Aharony:1997bx}. In the 
duality between M- and F-theory, it is crucial to include all Kaluza-Klein (KK) states in the loop 
\cite{Cvetic:2012xn,Cvetic:2013uta},\footnote{See also \cite{Grimm:2013oga} for the case of 
CS-terms in 5D M-/F-theory duality.} yielding the full loop corrected  CS-terms 
expression
\beq \label{eq:thetaF}
	\Theta^F_{AB}=
\Theta^F_{\text{cl},\, AB}
+ \frac12\sum_{\underline{q}}n(\underline{q})q_A q_B \ \text{sign}(q_A\zeta^A)\,.
\eeq
Here  $n(\underline{q})$ is the number of 3D fermions with charge vector $\underline{q}=(q_0,q_\alpha,q_i,q_m)$. It includes the charge  
$q_0$ w.r.t.~the 3D graviphoton, i.e.~the KK-level of states, the charges $q_\alpha$, $\alpha=1,\ldots,h^{(1,1)}(B)$,   under
3D vectors dual to the K\"ahler moduli of $B$, the charges $q_i$, $i=1,\ldots, \text{rk}(G)$, and $q_m$, $m=1,\ldots, r$, w.r.t.~to 4D Cartan gauge fields of
the non-Abelian gauge group $G$  of F-theory 
and the $r$ U(1) gauge fields, respectively.
The real parameters $\zeta^A$ are the Coulomb branch 
parameters.

Duality requires an identification of  
the CS-terms on the F-theory side with those in \eqref{eq:thetaM} on the M-theory side 
\cite{Grimm:2011fx,Grimm:2012rg,Cvetic:2012xn,Braun:2013yti,Cvetic:2013uta,Grimm:2015zea},
\beq \label{eq:M=F}
	\Theta_{AB}\equiv \Theta_{AB}^M\stackrel{!}{=}\Theta_{AB}^F\,.
\eeq
This immediately leads to additional restrictions on the CS-terms in F-theory 
\cite{Marsano:2011hv,Grimm:2011fx,Cvetic:2012xn,Cvetic:2013uta}, because certain CS-terms 
$\Theta_{AB}^F$ in F-theory
computed according to \eqref{eq:thetaF} are identically zero.
Physically, the implied constraints on the $G_4$-flux ensure the absence of circle flux in the circle 
compactification from F- to M-theory, an unbroken non-Abelian gauge group in 4D due to the absence 
of axion gaugings and the absence of non-geometric effects,
\begin{align}\label{eq:csfluxconstraints}
\Theta_{0\alpha}  = \Theta_{i\alpha} = \Theta_{\alpha \beta}= 0 \;.
\end{align}
Here we have to chose the basis $D_A$ of $H^{(1,1)}(X)$ so that index $0$ corresponds to the zero 
section $\hat{s}_0$ of the fibration of $X$, $\alpha=1,\ldots,h^{(1,1,)}(B)$, labels the vertical  
divisors induced  from the base $B$, $i=1,\ldots, \text{rk}(G)$ labels the Cartan divisors of $X$, 
where $G$ as before is non-Abelian part of the F-theory gauge group,  and $m=1,\ldots,r$ labels the 
$r$ U(1)-factors corresponding to  Shioda maps $\sigma(\hat{s}_m)$ 
of the rank $r$ Mordell-Weil (MW) group of rational sections $\hat{s}_m$ of $X$.

\paragraph{Chiralities in F-theory and $G_4$-flux quantization:} 

In order to calculate the matter chiralities $\chi(\mathbf{R})$ for a given matter 
representation $\mathbf{R}$ in a four-dimensional F-theory compactification,
we need to integrate the $G_4$-flux over a corresponding matter surface in $X$. 
The relevant matter surface $\mathcal{C}^w_\mathbf{R}$ 
is given as the rational surface constructed by fibering a $\mathbb{P}^1$ 
carrying the weight $\mathbf{w}$ of the representation $\mathbf{R}$ over the corresponding 
matter curve in the base $B$. The 4D chirality of $\mathbf{R}$ is computed as
\begin{align}\label{eq:g4chiralities}
\chi (\mathbf{R})&=n(\mathbf{R})-n(\bar{\mathbf{R}})=\int_{\mathcal{C}^w_\mathbf{R}} G_4 \; ,
\end{align}
where $n(\mathbf{R})$ denotes the number of left-chiral Weyl fermions in the representation 
$\mathbf{R}$.

Technically, the determination of the $\mathcal{C}^w_\mathbf{R}$ can be involved and requires the 
computation of the homology class of prime ideals describing the given matter surface. 
This can be done
using  the resultant technique that was applied first in \cite{Cvetic:2013nia,Cvetic:2013uta} for 
F-theory and will be exemplify for the three examples studied in this work.
As a consistency check of our geometric computations,  following 
\cite{Grimm:2011fx,Cvetic:2012xn,Cvetic:2013uta}, we use the matching condition 
\eqref{eq:M=F} of the CS-terms to double-check the 4D chiralities calculated using
\eqref{eq:g4chiralities}. 

Finally, let us comment on $G_4$-flux quantization.
In principal, in order to address $G_4$-flux quantization we have to expand $G_4$ and $c_2(X)$
in an integral basis for $H^{(2,2)}_V(X)$  and check the condition \eqref{eq:quantization}. This
integral basis can be determined employing mirror symmetry techniques 
\cite{Grimm:2009ef,Jockers:2009ti,Bizet:2014uua}. Since this is beyond the scope of 
this work, we will apply an indirect approach to ensure integral $G_4$-flux.

Here we exploit that $G_4$-flux quantization \eqref{eq:quantization}, the integrality of the
number $n_{\rm D3}$ of D3-branes, that is a necessary condition for quantized 
$G_4$-flux \cite{Witten:1996md}, the integrality of the CS-terms \eqref{eq:thetaM} and of the 
chiralities \eqref{eq:g4chiralities} are obviously linked to each other.
Thus, our strategy will be the following. First, we compute all chiralities $\chi(\mathbf{R})$ 
using \eqref{eq:g4chiralities}. Then, we parametrize the coefficients in the expansion of the $G_4$-flux 
w.r.t.~a basis of $H^{(2,2)}_V(X)$ in terms of these integral chiralities.
We then impose the necessary condition of integrality and positivity of $n_{\rm  D3}$. This will yield in 
turn constraints in form of lower bounds on the 4D chiralities. Next, we impose, if possible,
a family structure on our model.
Finally, we check that for this phenomenologically preferred 
choice of $G_4$-flux all CS-terms are integral, which ensures that the quantization condition \eqref{eq:quantization} is obeyed.

\paragraph{Toric hypersurface fibrations for 4D chiral F-theory models:} 

In order to introduce some notation used throughout this work, we conclude this introductory 
section with a very brief review of CY-fourfolds $X$
constructed as toric hypersurface fibrations. A detailed account on this subject can be found in
\cite{Klevers:2014bqa}.

We consider here elliptically fibered Calabi-Yau manifolds $X_{F_i}$ whose elliptic fiber is 
realized as the general CY-hypersurface in a 2D toric variety $\mathbb{P}_{F_i}$ associated
to one of the 2D reflexive polyhedra $F_i$. Here we focus on the polyhedra $F_{11}$, $F_{13}$ and 
$F_{16}$ in \cite{Klevers:2014bqa}, that naturally yield phenomenologically interesting models. In 
these cases, the corresponding toric ambient varieties $\mathbb{P}_{F_i}$ of the elliptic fiber
are blow-ups of $\mathbb{P}^2$. The elliptic curves in all considered cases is consequently given as an 
appropriate specialization of the general cubic
\beq \label{eq:pF1}
p = s_1 u^3 + s_2 u^2 v+ s_3 u v^2 + s_4 v^3 + s_5 u^2 w + s_6 u v w + s_7 v^2 w + s_8 u w^2+s_9 v w^2 +s_{10} w^3\,.
\eeq
Here the coefficients $s_i$ take values in a field $K$ and $[u:v:w]$ are projective coordinates on 
$\mathbb{P}^2$.

An elliptic fibration $X_{F_i}$ with fiber given by \eqref{eq:pF1} or specializations thereof is 
constructed by first fibering  the toric ambient space $\mathbb{P}_{F_i}$ over a chosen base $B$, 
then imposing \eqref{eq:pF1} and finally 
demanding the CY-condition. In this procedure, the coordinates $[u:v:w]$ and the coefficients $s_i$ in 
\eqref{eq:pF1} are lifted to sections of appropriate line bundles on $B$. 
The CY-condition fixes these line bundles to the following:
\beq \label{eq:cubicsections}
\text{
\begin{tabular}{c|c}
\text{section} & \text{Line Bundle}\\
\hline
	$u$&$\mathcal{O}(H+\cS_9+[K_B])$\rule{0pt}{13pt} \\
	$v$&$\mathcal{O}(H+\cS_9-\cS_7)$\rule{0pt}{12pt} \\
	$w$&$\mathcal{O}(H)$\rule{0pt}{12pt} \vspace{2cm}\\
\end{tabular}
}\qquad \text{
\begin{tabular}{c|c}
\text{section} & \text{Line Bundle}\\
\hline
	$s_1$&$\mathcal{O}_B(3[K_B^{-1}]-\cS_7-\cS_9)$\rule{0pt}{13pt} \\
	$s_2$&$\mathcal{O}_B(2[K_B^{-1}]-\cS_9)$\rule{0pt}{12pt} \\
	$s_3$&$\mathcal{O}_B([K_B^{-1}]+\cS_7-\cS_9)$\rule{0pt}{12pt} \\
	$s_4$&$\mathcal{O}_B(2\cS_7-\cS_9)$\rule{0pt}{12pt} \\
	$s_5$&$\mathcal{O}_B(2[K_B^{-1}]-\cS_7)$\rule{0pt}{12pt} \\
	$s_6$&$K_B^{-1}$\rule{0pt}{12pt} \\
	$s_7$&$\mathcal{O}_B(\cS_7)$\rule{0pt}{12pt} \\
	$s_8$&$\mathcal{O}_B([K_B^{-1}]+\cS_9-\cS_7)$\rule{0pt}{12pt} \\
	$s_9$&$\mathcal{O}_B(\cS_9)$ \rule{0pt}{12pt} \\
	$s_{10}$&$\mathcal{O}_B(2\cS_9-\cS_7)$ \rule{0pt}{12pt}
\end{tabular}
}
\eeq
Here, $\mathcal{O}(D)$ denotes the line bundle associated to a divisor $D$,\footnote{A subscript
indicates the space over which this line bundle is defined, e.g.~$\mathcal{O}_B(D)$ denotes a line 
bundle over $B$. If a subscript is omitted, the line bundle lives on the ambient space of $X$.} $H$ is 
the hyperplane on $\mathbb{P}^2$, $[K_B^{-1}]$ is the anti-canonical divisor of $B$ and 
$\mathcal{S}_7$, $\mathcal{S}_9$ are the divisor classes of $s_7$, $s_9$, respectively. We note 
that the table on the right hand side in \eqref{eq:cubicsections} applies for all examples studied below.

\section{Minimal Supersymmetric Standard Model: \\$G_{F_{11}}=\text{SU(3)}\times\text{SU(2)}\times\text{U(1)}$}
\label{sec:standardmodel}

In this section we discuss an F-theory compactification on the elliptically fibered CY-manifold
$X_{F_{11}}$ which yields precisely the gauge group and representation content of the
Minimal Supersymmetric Standard Model (MSSM) \cite{Klevers:2014bqa}. 

In Section \ref{sec:geomF11} we elaborate on the basic geometrical properties of $X_{F_{11}}$ that 
encode the gauge symmetry, including the $\U1$ generator, as well as the matter representations. 
While these observations are model independent, we further specialize to the simple base 
$B=\mathbb{P}^3$.  For this specific case we compute the vertical cohomology 
$H^{(2,2)}_V(X_{F_{11}})$ in Section \ref{sec:G4F11}. Using these results, we explicitly construct 
$G_4$-flux consistent with all F-theory consistency constraints. We compute the induced 4D chiralities 
of the matter representations, that we double-check employing 3D CS-terms and M-/F-theory duality. 
Next in Section \ref{sec:anomalien} we discuss 4D anomaly cancelation and the properties of models 
which exhibit a complete family structure, in particular the existence of three family models with 
positive and integral D3-brane charge and quantized $G_4$-flux. In Section \ref{sec:phenoF11} we 
conclude with some comments on the phenomenology of the three family models we found.

The elliptic fibration $X_{F_{11}}$ has been completely analyzed in \cite{Klevers:2014bqa},
to which we refer for more details on its codimension one, two and three singularities
and the corresponding 6D F-theory compactification. The relevant results are summarized in Section 
\ref{sec:geomF11}. The reader less interested in the technical details can directly jump to the
4D chiralities in \eqref{eq:f11_chiralities} and the following discussions.

\subsection{The Geometry of Gauge Symmetry and Particle Representations}
\label{sec:geomF11}
\begin{figure}[H]
\centering
\begin{minipage}{.56\textwidth}
  \centering
  \includegraphics[scale=.4]{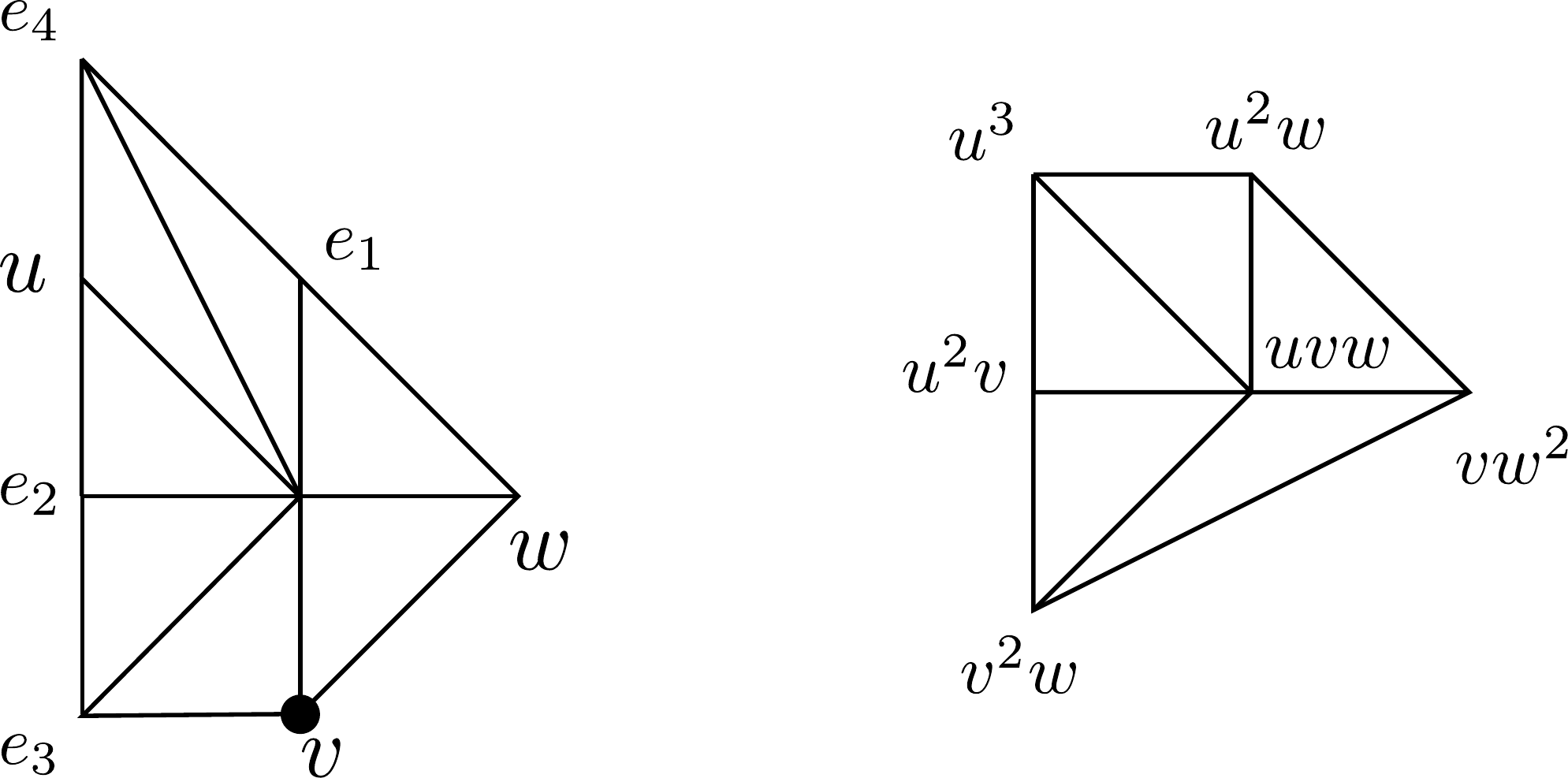}
\end{minipage}%
\begin{minipage}{.44\textwidth}
{\footnotesize
  \begin{tabular}{|c|c|}\hline
Section & Line Bundle\\ \hline
$u$ & $\mathcal{O}(H-E_1-E_2-E_4+\cS_9+[K_B])$ \\ \hline
$v$ & $\mathcal{O}(H-E_2-E_3+\cS_9-\cS_7)$\\ \hline
$w$ & $\mathcal{O}(H-E_1)$\\ \hline
$e_1$ & $\mathcal{O}(E_1-E_4)$\\ \hline
$e_2$ & $\mathcal{O}(E_2-E_3)$\\ \hline
$e_3$ & $\mathcal{O}(E_3)$\\ \hline
$e_4$ & $\mathcal{O}(E_4)$\\ \hline
\end{tabular}}
\end{minipage}
\caption{\label{fig:poly11_toric}The toric diagram of polyhedron $F_{11}$ and its dual. The zero section is indicated by the dot. In the accompanying table we indicate the divisor classes of the fiber coordinates.}
\end{figure}
The elliptic fiber which is used to engineer F-theory models that naturally exhibit the gauge symmetry of the standard model is 
given as the CY-hypersurface
\begin{align}
\begin{split}\label{eq:pF11}
p_{F_{11}}&=s_1 e_1^2 e_2^2 e_3 e_4^4 u^3 + s_2 e_1 e_2^2 e_3^2 e_4^2 u^2 v + s_3 e_2^2 e_3^2 u v^2 + s_5 e_1^2 e_2 e_4^3 u^2 w 
+s_6 e_1 e_2 e_3 e_4 u v w + s_9 e_1 v w^2 \,
\end{split}
\end{align}
in the toric ambient space $\mathbb{P}_{F_{11}}$. Its toric data is summarized in Figure
\ref{fig:poly11_toric}. The divisor classes in $\mathbb{P}_{F_{11}}$ are $H$, the hyperplane class of
$\mathbb{P}^2$, as well as the four exceptional divisors $E_1$, $E_2$, $E_3$ and $E_4$.

Next, an elliptically fibered CY-fourfold $X_{F_{11}}$ with the  elliptic fiber \eqref{eq:pF11} is constructed by promoting
the coefficients $s_i$ in the CY-equation to sections of the line bundles of $B$ given in \eqref{eq:cubicsections}. The elliptic 
fibration of $X_{F_{11}}$ is equipped with two independent rational sections
\begin{align}
\begin{split}
\hat{s}_0=X_{F_{11}}\cap\{v=0\}\,\,&:\quad [1:0:s_1:1:1:-s_5:1]\,,\\
\hat{s}_1=X_{F_{11}}\cap\{e_4=0\}&:\quad[s_9:1:1:-s_3:1:1:0]\,,
\end{split}
\end{align}
where we have chosen $\hat{s}_0$ as the zero section.
By computation of the discriminant of the fibration \eqref{eq:pF11}, one can check that over the loci
$\cS_{\SU2}=\{s_3=0\}$ and $\cS_{\SU3}=\{s_9=0\}$ the fiber degenerates to
$I_2$- and $I_3$-fibers giving rise to $\SU2$ and $\SU3$ gauge symmetries, respectively. The Cartan divisors of these gauge groups are
\begin{align}
D^{\SU2}_{1}=[e_1] \, ,\quad
D^{\SU3}_{1}=[e_2] \, \quad
D^{\SU3}_{2}=[u] \,.
\end{align}
Having these divisors at hand, one can show that the generator of the \U1 symmetry, that is
the Shioda  map of $\hat{s}_1$, is given by
\begin{align} \label{eq:ShiodaF11}
\sigma (\hat{s}_1)=S_1-\tilde{S}_0 +  [K_B] + \frac{1}{2} D^{\SU2}_{1}	+ \frac{1}{3}\left(D^{\SU3}_{1}+2D^{\SU3}_{2}\right) \, .
\end{align}
Here, $S_1$ denotes the class of $\hat{s}_1$ and we used
$\tilde{S}_0=S_0+\tfrac12 [K_B^{-1}]$ \cite{Bonetti:2011mw}, where $S_0$ is the class of 
$\hat{s}_0$ and $K_B^{-1}$ denotes the anti-canonical bundle of the base $B$.
The corresponding N\'{e}ron-Tate height pairing reads
\begin{align}
b_{11}=\frac32[K_B^{-1}] - \frac12\cS_7 - \frac16\cS_9\; .
\label{eq:b11f11}
\end{align}

Furthermore, there are codimension two singularities in the elliptic fibration $X_{F_{11}}$
which support all matter representations of the Standard Model\footnote{At this stage, over the different codimension two loci, matter comes in
vector-like pairs. It is the $G_4$-flux that induces chiralities for the fields.} as one can see in Table
\ref{tab:poly11_matter}.

\begin{table}[h]
\begin{center}
\renewcommand{\arraystretch}{1.2}
\begin{tabular}{|c|@{}c@{}|}\hline
Representation & Locus \\ \hline
$(\three,\two)_{1/6}$ & $V(I_{(1)}):=\{ s_3=s_9=0 \}$ \\ \hline
$(\one,\two)_{-1/2}$ & $ V(I_{(2)}):=\{ s_3=s_2s_5^2 + s_1(s_1 s_9-s_5s_6)=0 \}$ \\ \hline
$(\overline{\three},\one)_{-2/3}$ & $V(I_{(3)}):=\{ s_5=s_9=0 \}$ \\ \hline
$(\overline{\three}, \one)_{1/3}$ & $  V(I_{(4)}):=\{ s_9= s_3 s_5^2 + s_6(s_1 s_6-s_2 s_5)=0 \}$ \\ \hline
$(\one,\one)_{1}$ & $V(I_{(5)}):=\{ s_1=s_5=0 \}$ \\ \hline
\end{tabular}
\caption{\label{tab:poly11_matter} Charged matter representations under $\SU3 \times \SU2 \times \text{U}(1)$ and corresponding codimension two loci in $X_{F_{11}}$. The charge under the $\U1_Y$ generator is indicated by a subscript.}
\end{center}
\end{table}

We note that the second Chern class $c_2$ and the Euler number of $X_{F_{11}}$ can be computed
base independently \cite{Cvetic:2013qsa}. They are needed to check  the $G_4$-flux
quantization condition \eqref{eq:quantization} as well as the cancelation of D3 tadpoles
\eqref{eq:tadpole}. We obtain
\begin{align}
  c_2(X_{F_{11}})&= -c_1^2 + c_2 + c_1 E_2 - c_1 E_3 - 7 c_1 E_4 - 7 E_4^2 + 4 c_1 H + 2 c_1 \cS_7 + 4 E_1 \cS_7 + E_2 \cS_7 \\
		&\phantom{=}+ E_3 \cS_7 + 6 E_4 \cS_7 - 4 H \cS_7 - c_1 \cS_9 - 5 E_1 \cS_9 - 3 E_2 \cS_9 - E_3 \cS_9 + 3 H \cS_9 - 3 \cS_7 \cS_9 + 3 \cS_9^2\; ,\nonumber\\
  \chi(X_{F_{11}})&=3 (24 c_1^3+4 c_1 c_2-16 c_1^2 \cS_7+8 c_1 \cS_7^2-18 c_1^2 \cS_9+3 c_1 \cS_7 \cS_9-3 \cS_7^2 \cS_9+6 c_1 \cS_9^2+\cS_7 \cS_9^2) \label{eq:eulerf11} \; ,
\end{align}
where $c_1$ and $c_2$ denote the first and second Chern class of the base $B$, respectively. The 
divisors $\mathcal{S}_7$ and $\mathcal{S}_9$ are introduced in \eqref{eq:cubicsections}. 

\begin{figure}[H]
\center
\includegraphics[scale=0.8]{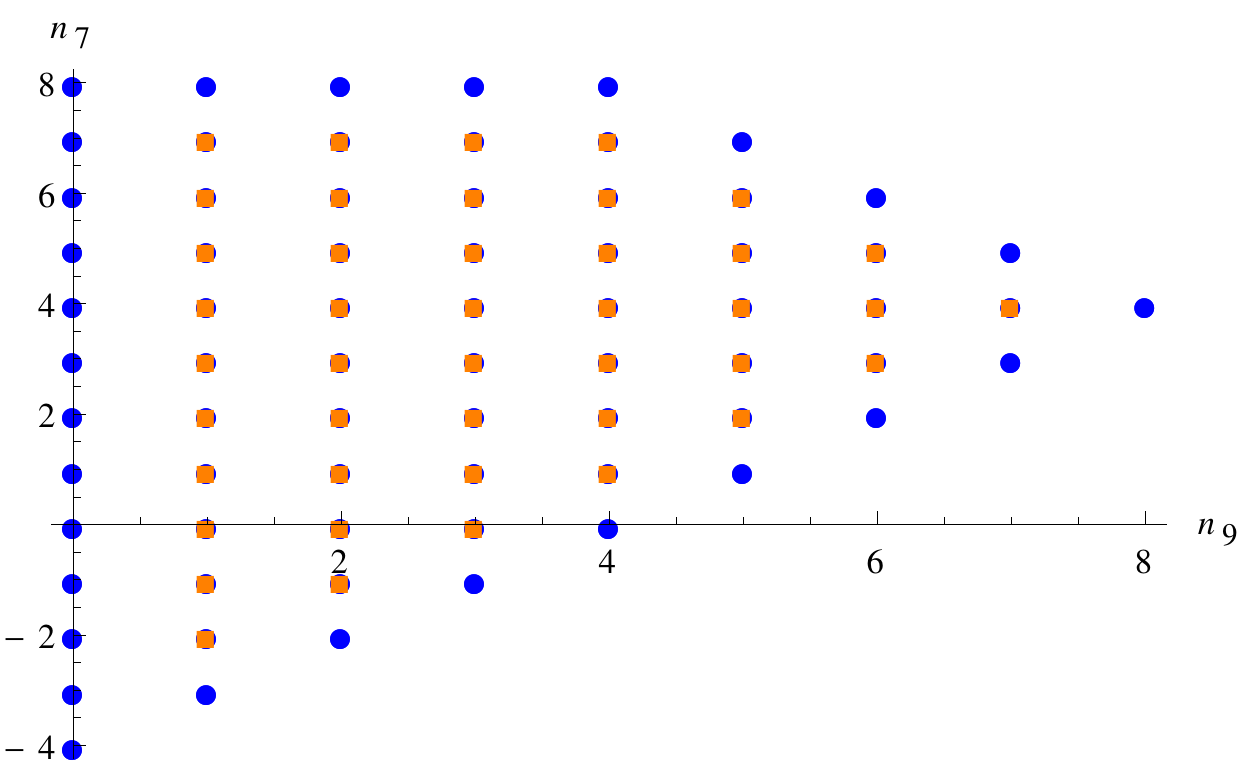}
\caption{\label{fig:f11allowed} Allowed region for $(n_7,n_9)$ for the  CY-fourfold $X_{F_{11}}$ with 
base $\mathbb{P}^3$. Orange dots indicate that all SM representations are present and a
$G_4$-flux admitting $b$  families exists.}
\end{figure}

For the remainder of this section, we chose the base of the fibration to be $B=\mathbb{P}^3$.  For this
simple choice of base the only  vertical divisor is the pullback of the hyperplane class
$\mathbb{P}^3$, which we denote by $H_B$. In this case, we have $c_1=4[H_B]$ and 
$c_2=6[H_B^2]$ in \eqref{eq:eulerf11}. In addition,  we readily expand the divisors $\mathcal{S}_7$,
$\mathcal{S}_9$ in \eqref{eq:cubicsections} needed to specify the fibration $X_{F_{11}}$ in terms of
$H_B$,
\begin{align}\label{eq:baseP3}
\cS_7 = n_7 H_B\; , \quad
\cS_9 = n_9 H_B\; , \quad
[K_B^{-1}]&=4 H_B\,,
\end{align}
where $n_7$ and $n_9$ denote integers. These integers are constrained by requiring
effectiveness of all divisor classes in \eqref{eq:cubicsections}, that enter the CY-constraint 
\eqref{eq:pF11}. This determines a region of allowed values for
the pair $(n_7,n_9)$, to which we refer to as the allowed region, as depicted in Figure \ref{fig:f11allowed}.

\subsection{$G_4$-Flux and Matter Chiralities}\label{sec:G4F11}

For the specific base $B=\mathbb{P}^3$, the full SR-ideal of the toric ambient space
of $X_{F_{11}}$ is given by\footnote{The SR-ideal of the fiber alone can be found in \cite{Klevers:2014bqa}.}
\begin{align}\label{eq:SRF11}
SR_{F_{11}}&=\{ u e_1, uw, uv, ue_3, e_4 w, e_4 v, e_4 e_3, e_4 e_2, e_1v, e_1 e_3, e_1e_2, w e_3, w e_2, ve_2, x_0x_1x_2x_3\}
\end{align}
where $[u,v,w]$ and the $e_i$, $i=1,\ldots,4$, are the projective coordinates on the fiber and $x_j$ 
($j=0,1,2,3$) are the homogeneous coordinates on the $\mathbb{P}^3$ base. A basis for
$H^{(1,1)}(X_{F_{11}})$ is given by
\begin{align}
H^{(1,1)}(X_{F_{11}})&=\langle H_{B}, \tilde{S}_0, D^{\SU2}_{1}, D^{\SU3}_{1}, D^{\SU3}_{2}, \sigma(\hat{s}_1) \rangle \, ,
\end{align}
where we denote, by abuse of notation, divisors and their Poincare dual $(1,1)$-forms by the same 
symbol.

Next we proceed with the computation of the full vertical cohomology ring of $X_{F_{11}}$ following
\cite{Cvetic:2013uta}. To set up its computation as a quotient ring, we need the SR ideal
\eqref{eq:SRF11}, the basis of divisors \eqref{eq:baseP3} as well as the intersection numbers,
\beq
 H_B^3 \cdot S_0^2=-1\,, \qquad  H_B^3\cdot S_1^2=-1\,,
\eeq
which follow from the toric intersections in $\mathbb{P}_{F_{11}}$.
The quartic intersections of $X_{F_{11}}$ can be readily computed and a canonical basis for
$H^{(3,3)}(X_{F_{11}})$ is obtained by duality to $H^{(1,1)}(X_{F_{11}})$.
We obtain generators for $H^{(2,2)}_V(X_{F_{11}})$ by constructing all possible products of two
divisors in $H^{(1,1)}(X_{F_{11}})$. We evaluate the rank of the inner product on these generators as
\begin{align}
{\rm dim} H^{2,2}(X_{F_{11}}) &= 7 \; .
\end{align}
As a  basis for $H^{(2,2)}_V(X_{F_{11}})$ we choose the seven elements
\begin{align}
H^{(2,2)}_V(X_{F_{11}})&=\langle (H_{B})^2,\; H_{B} \tilde{S}_0,\; D^{\SU2}_{1} H_{B},\; D^{\SU3}_{1} H_{B},\; D^{\SU3}_{2} H_{B},\; H_{B} \sigma(\hat{s}_1),\; \tilde{S}_0^2 \rangle\;.
\end{align}

We can then expand the $G_4$-flux in terms of the this basis.  Imposing the conditions
\eqref{eq:csfluxconstraints} required by a match of M- and F-theory CS-terms leads to five conditions
on the $G_4$-flux, yielding the following two parameter $G_4$-flux on $X_{F_{11}}$:
\begin{align}
\begin{split}\label{eq:fluxf11}
G_4&= a_6H_{B} \cdot\sigma(\hat{s}_1) -a_7\left[\tilde{S}_0^2+ (20 n_7-n_7^2+8 n_9-n_7 n_9-92)H_{B}^2 \right]\;
\end{split}
\end{align}
Here $a_6$ and $a_7$ are free discrete parameters entering the $G_4$-flux. Their quantization is
fixed by the $G_4$-flux quantization condition \eqref{eq:quantization} using the
the expression \eqref{eq:eulerf11} for $c_2(X_{F_{11}})$. Solving flux quantization in general
requires the knowledge of the integral basis for $H^{(2,2)}_V(X_{F_{11}})$. Since the determination
of the general integral basis is beyond the scope of this work, we will check $G_4$-flux quantization in
dependence on the number of chiral families indirectly by ensuring an integral and positive number
$n_{\rm D3}$ of D3-branes and quantization of the 3D CS-terms \eqref{eq:thetaM}. For the detailed 
discussion, we refer to Section \ref{sec:anomalien}.

In order to compute the 4D matter chiralities we have to compute the homology classes of the
matter surfaces for all representations in Table \ref{tab:poly11_matter}. For those codimension two
matter surfaces given as complete intersections in the toric ambient space of $X_{F_{11}}$,
the homology classes of the corresponding matter
surfaces follow directly from the second column in Table \ref{tab:poly11_matter} and the splitting of the
fiber \eqref{eq:pF11} at the respective locus, cf.~\cite{Klevers:2014bqa}. We obtain
\begin{align} \label{eq:F11MatS1}
\begin{split}
\mathcal{C}^w_{(\three,\two)_{1/6}}&=	\cS_9\cdot ([K_B^{-1}] + \cS_7 - \cS_9)\cdot E_4\;,\\
\mathcal{C}^w_{(\overline{\three},\one)_{-2/3}}&=	-\cS_9\cdot (2 [K_B^{-1}] - \cS_7) \cdot(2 H - E_2 - E_3 - \cS_7 + \cS_9 + [K_B^{-1}])\;,\\
\mathcal{C}^w_{(\one,\one)_{1}}\,\,&=	-(2 [K_B^{-1}] - \cS_7)\cdot (3 [K_B^{-1}] - \cS_7 - \cS_9)\cdot (2 H - E_1 - E_4 + \cS_9)\;,
\end{split}
\end{align}
where we have used \eqref{eq:cubicsections}. Here, we have chosen a node in the respective fiber at
codimension two that is not intersected by the zero section.
The matter surfaces for the two matter loci supporting the representations $(\one,\two)_{-1/2}$ and
$(\overline{\three},\one)_{1/3}$ are not complete intersections. Their associated prime ideals are computed by
a primary decomposition. Their respective  homology classes are obtained by choosing a suitable
complete intersection containing a given matter surface and by subtracting all its other irreducible
components with their corresponding multiplicities as determined by the resultant. We obtain
\begin{align}\label{eq:F11MatS2}
\begin{split}
[\mathcal{C}^w_{(\one,\two)_{-1/2}}]&= \big[(D^{\SU2}_{1}+D^{\SU3}_{1}+2 D^{\SU3}_{2}+4 [K_{B}^{-1}]+3 S_1-2 \cS_7)\cdot (6 [K_{B}^{-1}]-2 \cS_7-\cS_9) \\
  &\phantom{=} -2 (2 [K_{B}^{-1}]-\cS_7) \cdot(3 [K_{B}^{-1}]-\cS_7-\cS_9)+2 (-2 [K_{B}^{-1}]+\cS_7)\cdot \cS_9\big]\cdot (\cS_9-\cS_7-[K_{B}^{-1}])\; , \\
[\mathcal{C}^w_{(\overline{\three}, \one)_{1/3}}]&= \big[2 [K_{B}^{-1}] \cdot(\cS_7-2 [K_{B}^{-1}])+(D^{\SU2}_{1}+D^{\SU3}_{2}+2 [K_{B}^{-1}]+2 S_1-\cS_7) \\
  &\phantom{=} \times (5 [K_{B}^{-1}]-\cS_7-\cS_9)\big]\cdot \cS_9 \; .
\end{split}
\end{align}
Finally,  we compute the integrals \eqref{eq:g4chiralities} of the  $G_4$-flux in \eqref{eq:fluxf11} over
the various matter surfaces in \eqref{eq:F11MatS1} and \eqref{eq:F11MatS2}, yielding the following 4D
matter chiralities:
\begin{align}\begin{split}\label{eq:f11_chiralities}
\chi_{(\three,\two)_{1/6}}&=	\tfrac16 (4+n_7-n_9) n_9 a_6				 \, ,			\\
\chi_{(\one,\two)_{-1/2}}\,\,\,&=	\tfrac12 (4+n_7-n_9) ((2 n_7+n_9-24) a_6+4 (n_7-8) (n_7+n_9) a_7-12)	\\
\chi_{(\overline{\three},\one)_{-2/3}}&=	\tfrac13 (n_7-8) n_9 (2 a_6+3 (n_7+n_9-12) a_7)		\, ,		\\
\chi_{(\overline{\three}, \one)_{1/3}}\,\,\,&=	-\tfrac13 n_9 ((n_7+n_9-20) a_6+3 (n_7-8) (n_7+n_9-12) a_7)	\, ,		\\
\chi_{(\one,\one)_{1}}\,\,\,\,&=	(n_7-8) (n_7+n_9-12) (a_6+(2 n_7+n_9-16) a_7)				\; .
\end{split}
\end{align}

As a cross-check of these geometric results, we use the duality between M- and F-theory in three dimensions and the implied matching of 3D CS-terms
\cite{Grimm:2011fx,Cvetic:2012xn,Cvetic:2013uta} to compute the 4D chiralities.
For the spectrum in Table \eqref{tab:poly11_matter}, we  compute all nonzero CS-terms on the
F-theory side as
\bea
\Theta^\mathrm{F}_{m=1,n=1} \!\!&\!=\!&\!\! \tfrac12 (\tfrac16 \chi_{(\three,\two)_{1/6}} - \tfrac12 \chi_{(\one,\two)_{-1/2}} - \tfrac43 \chi_{(\overline{\three},\one)_{-2/3}} + \tfrac13 \chi_{(\overline{\three}, \one)_{1/3}} + 3 \chi_{(\one,\one)_{1}})\;,\\
\Theta^\mathrm{F}_{i=2,j=2} \!\!&\!=\!&\!\! 3 \chi_{(\three,\two)_{1/6}} -\chi_{(\one,\two)_{-1/2}}\;,\qquad\qquad\qquad \quad\,\,\,\,
\Theta^\mathrm{F}_{i=3,j=3} = 2 \chi_{(\three,\two)_{1/6}} -\chi_{(\overline{\three},\one)_{-2/3}} + \chi_{(\overline{\three}, \one)_{1/3}}\, ,\nn\\
\Theta^\mathrm{F}_{i=3,j=4} \!\!&\!=\!&\!\! -\tfrac12 (2 \chi_{(\three,\two)_{1/6}} - \chi_{(\overline{\three},\one)_{-2/3}} + \chi_{(\overline{\three}, \one)_{1/3}})\; ,\quad	 \Theta^\mathrm{F}_{i=4,j=4} = 2 \chi_{(\three,\two)_{1/6}} - \chi_{(\overline{\three},\one)_{-2/3}} + \chi_{(\three, \one)_{1/3}}\, ,\nn
\eea
where the label $m=1$ corresponds to the divisor of the 4D U(1) gauge fields, $i=2$ to the Cartan
divisor $D^{\SU2}_{1}$ and $i=3,4$ to the Cartan divisors $D^{\SU3}_{1}$ and $D^{\SU3}_{2}$,
respectively. We note that only the SM-singlet has a non-trivial KK-charge, cf.~\cite{Klevers:2014bqa}.
We readily compute the CS-terms \eqref{eq:thetaM} on the M-theory side for the $G_4$-flux
\eqref{eq:fluxf11}. The matching $\Theta^\mathrm{F}_{AB}\stackrel{!}{=}\Theta^\mathrm{M}_{AB}$
precisely reproduces the chiralities in \eqref{eq:f11_chiralities}.

\subsection{4D Anomaly Cancelation and Family Structure}
\label{sec:anomalien}

As an additional cross-check of our computations, we can verify that all 4D anomalies are
canceled by a generalized Green-Schwarz mechanism \cite{Green:1984sg,Sagnotti:1992qw}.

For $X_{F_{11}}$, we have the following conditions implied by cancelation of the purely
non-Abelian, mixed Abelian-non-Abelian, purely Abelian and mixed Abelian-gravitational anomalies:
\begin{align}
\begin{split}
\label{eq:F11anomalies}
\SU3^3\,\,&:\,\, -2\chi_{(\three,\two)_{1/6}}+\chi_{(\overline{\three},\one)_{-2/3}}+\chi_{(\overline{\three},\one)_{1/3}}=0\,,\\
 \SU2^2-\U1\,\,&:\,\, {\textstyle \frac12}\left[3\chi_{(\three,\two)_{1/6}}{\textstyle\left(\frac16\right)}+\chi_{(\one,\two)_{-1/2}}{\textstyle\left(-\frac12\right)}\right]=-{\textstyle \frac18}b^{\alpha}_{\SU2}\Theta_{\alpha,m=1}\,,\\
 \SU3^2-\U1\,\,&:\,\, {\textstyle \frac12}\left[2\chi_{(\three,\two)_{1/6}}{\textstyle\left(\frac16\right)}+\chi_{(\overline{\three},\one)_{-2/3}}{\textstyle\left(-\frac23\right)}+\chi_{(\overline{\three},\one)_{1/3}}{\textstyle\left(\frac13\right)}\right]=-{\textstyle \frac18}b^{\alpha}_{\SU3}\Theta_{\alpha,m=1}\,,\\
\U1^3\,\,&:\,\,{\textstyle \frac16}\left[ 6\chi_{(\three,\two)_{1/6}}{\textstyle\left(\frac16\right)}^3+2\chi_{(\one,\two)_{-1/2}}{\textstyle\left(-\frac12\right)}^3\right.\\
&\,\,\quad\left. +3\chi_{(\overline{\three},\one)_{-2/3}}{\textstyle\left(-\frac23\right)}^3+3\chi_{(\overline{\three},\one)_{1/3}}{\textstyle\left(\frac13\right)}^3+\chi_{(\one,\one)_{1}}(1)^3\right]=-{\textstyle \frac18}b^{\alpha}_{11}\Theta_{\alpha,m=1}\,,\\
{\rm Grav.}^2-\U1\,\,&:\,\, {\textstyle \frac{1}{48}}\left[6\chi_{(\three,\two)_{1/6}}{\textstyle\left(\frac16\right)}+2\chi_{(\one,\two)_{-1/2}}{\textstyle\left(-\frac12\right)}\right.\\
&\,\,\left.\quad+3\chi_{(\overline{\three},\one)_{-2/3}}{\textstyle\left(-\frac23\right)}+3\chi_{(\overline{\three},\one)_{1/3}}{\textstyle\left(\frac13\right)}+\chi_{(\one,\one)_{1}}(1)\right]={\textstyle\frac{1}{32}}a^{\alpha}\Theta_{\alpha,m=1}\,.
\end{split}
\end{align}
We recall that the index $\alpha$ runs over base divisors. For $\mathbb{P}^3$ we only have
$\alpha=1$ for the single vertical divisor $H_B$. The coefficients $b_{11}^{\alpha=1}$,
$b_{\SU2}^{\alpha=1}$ and $b_{\SU3}^{\alpha=1}$ can be computed as intersections of  $H_B^2$
with the N\'{e}ron-Tate height pairing $b_{11}$ \eqref{eq:b11f11}, and the GUT divisors $S_{\SU2}$,
$S_{\SU3}$, respectively. They read, written in terms of the integers $n_7$, $n_9$ introduced in
\eqref{eq:baseP3}, as
\begin{align}
b_{11}^{\alpha=1}=6 - \frac12 n_7 - \frac16 n_9\,,\quad b_{\SU2}^{\alpha=1}=4+n_7-n_9\,,\quad b_{\SU3}^{\alpha=1}=n_9\,.
\end{align}
The coefficient $a^\alpha$ in \eqref{eq:F11anomalies} appearing in the mixed Abelian-gravitational
anomaly stems from expanding $K_B$ in terms of the vertical divisors. For this particular case we have
$a^{\alpha=1}=-4$, according to \eqref{eq:baseP3}. Finally, the CS-term $\Theta_{\alpha=1,m=1}$
for the axion gauging is given by
\begin{align} \label{eq:gaugingXF11}
\Theta_{\alpha=1,m=1}={\textstyle \frac{1}{6}}\left[(-36+3 n_7+n_9) a_6+6 (-8+n_7) (-12+n_7+n_9) a_7\right]\,,
\end{align}
as we compute using the $G_4$-flux in \eqref{eq:fluxf11} and the general formula \eqref{eq:thetaM}.
Using these results and the chiralities \eqref{eq:f11_chiralities}, we find that all anomalies in relations
\eqref{eq:F11anomalies} are indeed satisfied.

Regarding the anomalies there are two remarks in order. First, we recall that even though the pure \SU2
anomaly is trivial, one has to guarantee that the model does not have a Witten anomaly, i.e.~the
spectrum of the theory must always exhibit an even number of doublets \cite{Witten:1982fp}. In our case, the Witten anomaly takes the form
\begin{equation} \label{eq:WittenSU(2)}
3\chi_{(\three,\two)_{1/6}}+\chi_{(\one,\two)_{-1/2}}\in 2\mathbb{Z}\,
\end{equation}
Using again the expressions given in \eqref{eq:f11_chiralities}, we see that the Witten anomaly is
canceled only if $a_6(n_7+n_9)$ is an even number. This is in contrast to the anomalies
\eqref{eq:F11anomalies}, which are canceled independently of the specific values for $n_7$, $n_9$,
$a_6$ and $a_7$. We expect that \eqref{eq:WittenSU(2)} is automatically obeyed for appropriately
quantized $G_4$-flux. For a model exhibiting a family structure, which is the case of interest in the
following, we have $\chi_{(\three,\two)_{1/6}}=\chi_{(\one,\two)_{-1/2}}$ for which
\eqref{eq:WittenSU(2)} is trivially satisfied.

Second, for general $G_4$-flux, the axion gauging \eqref{eq:gaugingXF11} is non-zero, as required by
anomalies. This induces a mass-term for the \U1 gauge field of $X_{F_{11}}$. As the \U1 in this model
corresponds to the hypercharge $\text{U}(1)_{\text{Y}}$, we have to impose that it is massless for
the sake of the phenomenology of our model. For this reason we have to require
$\Theta_{\alpha=1,m=1}=0$,
which reduces \eqref{eq:fluxf11} to a one-parameter $G_4$-flux. In
the absence of the axion gauging all anomalies in \eqref{eq:F11anomalies} must vanish identically.
Precisely as in the Standard Model, the cancelation of all anomalies can only be achieved if the
chiralities for all fields coincide,~i.e. if the matter fields come in a certain number $b$ of
complete families. Written in terms of this parameter $b\equiv \chi_{(\three,\two)_{1/6}}$, the 
coefficients $a_6$ and $a_7$ take the following form:
\beq\label{eq:familyparameterf11}
a_6 = \frac{6 b }{\left(4+n_7 -n_9   \right)n_9} \, , \qquad
a_7 =  -\frac{b\left( 3 n_7 + n_9 -36  \right)}{n_9 \left(n_7 -8  \right) \left(4+n_7 - n_9   \right)\left(n_7 + n_9-12  \right)} \; .
\eeq
We note that the vanishing of the factors appearing in the denominator of the above equations define
the boundaries of the allowed region for $(n_7,n_9)$, see Figure \ref{fig:f11allowed}. The boundary
region has to be excluded to begin with since there one or more chiralities in \eqref{eq:f11_chiralities}
vanish, leading to a model unsuited for phenomenological applications.

Using \eqref{eq:familyparameterf11}, we can write the $G_4$-flux \eqref{eq:fluxf11} as a function of
$b$, $n_7$ and $n_9$. In this parametrization, we check, for every allowed value for $(n_7,n_9)$,
whether there are integral values of $b$ for which
the number $n_{\rm D3}$ of D3-branes needed to cancel the tadpole \eqref{eq:tadpole} is a positive
integer, as expected for a smooth CY-fourfold $X_{F_{11}}$ and appropriately quantized $G_4$-flux
\cite{Witten:1996md}. Additionally, we impose that all CS-terms \eqref{eq:thetaM} are integral, which
is equivalent to $G_4$-flux quantization as discussed in Section \ref{sec:review}.  Without adding 
additional horizontal $G_4$-flux,  these two conditions  impose the lower bounds on the number $b$ of
families shown in Table \ref{tab:families_f11} for all values of $(n_7,n_9)$ in the allowed region and  
together with the corresponding numbers  $n_{\rm D3}$ of D3-branes. 
Remarkably, this simple analysis shows that the
minimal value of generations $b$ obeying these constraints is three. We find $b=3$ generations with
$n_{\text{D3}}=64$ and  $n_{\text{D3}}=46$  for the two strata with $(n_7,n_9)=(2,5)$ and
$(n_7,n_9)=(5,6)$, respectively.\footnote{Adding horizontal $G_4$-flux can lower the number of 
D3-branes further.}
\begin{table}[ht!]
\begin{center}
 \renewcommand{\arraystretch}{1.4}
{\footnotesize
\begin{tabular}{c|ccccccc}
 {\large $_{n_7} \backslash ^{n_9}$} & 1 & 2 & 3 & 4 & 5 & 6 & 7 \\ \hline
 7 & {\huge -} & $( 27;16 )$ & {\huge -} & {\huge -} &  &  &\\
 6 & {\huge -} & $( 12;81 )$ & $( 21;42 )$ & {\huge -} & {\huge -} &  &  \\
 5 & {\huge -} & {\huge -} & $( 12;57 )$ & $( 30;8 )$ & {\huge -} & {\color{red}$( 3;46 )$} &  \\
 4 & $( 42;4 )$ & {\huge -} & $( 30;32 )$ & {\huge -} & {\huge -} & {\huge -} & {\huge -} \\
 3 & {\huge -} & $( 21;72 )$ & {\huge -} & {\huge -} & {\huge -} & $( 15;30 )$ &  \\
2 & $( 45;16 )$ & $( 24;79 )$ & $( 21;66 )$ & $( 24;44 )$ & {\color{red} $( 3;64 )$} &  &  \\
 1 & {\huge -} & {\huge -} & {\huge -} & {\huge -} &  &  &  \\
  0 & {\huge -} & {\huge -} & $( 12;112 )$ &  &  &  &  \\
 -1 & $( 36;91 )$ & $( 33;74 )$ &  &  &  &  &  \\
 -2 & {\huge -} &  &  &  &  &  &  \\
\end{tabular}
}
\caption{\label{tab:families_f11}The entries $(b,n_{\text{D3}})$ show  the minimal number of families
$b$ for which the number $n_{\text{D3}}$ of D3-branes  is integral and positive for integral 3D
CS-terms. At the allowed points for $(n_7, n_9)$ marked as "-" the number of D3-branes is negative
for all positive integral values of $b$.}
\end{center}
\end{table}
\subsection{Phenomenological Discussion}
\label{sec:phenoF11}

The discussion in the previous section shows that only the two
models with $(n_7,n_9)=(2,5),\,(5,6)$ admit three chiral families, cf.~Table \ref{tab:families_f11}.
Note also that $b=3$ is the smallest permitted number of generations. Having found these three family solutions, we proceed in this section with the discussion of  the phenomenology of the model.

We begin by identifying  the representations from Table \ref{tab:poly11_matter} with the Standard
Model particles they correspond to:
\begin{align}
\label{tab:SM}
\text{
\begin{tabular}{|c|c|c|c|c|c|}\hline
$Q_i$ & $\bar{u}_i$ & $\bar{d}_i\rule{0pt}{13pt}$ & $L_i$ & $\bar{e}_i$ & $H_u$, $H_d$ \\ \hline
$(\three,\two)_{1/6}$ & $(\overline{\three},\one)_{-2/3}\rule{0pt}{13pt}$ & $(\overline{\three},\one)_{1/3}$ & $(\one,\two)_{-1/2}$ & $(\one,\one)_1$ & $(\one,\two)_{\pm 1/2}$ \\ \hline
\end{tabular}}
\end{align}
Here the index $i=1,2, 3$ labels the families and we use the common notation to denote quarks by
$Q_i$, $\bar{u}_i$ and $\bar{d}_{i}$,  leptons by $L_i$ and $\bar{e}_i$ and the two Higgses by
$H_u$ and $H_d$, respectively.

 In $X_{F_{11}}$ the Higgs fields
emerge as a vector-like pair from the same matter curve as the leptons $L_i$. In order to check
geometrically that there is indeed a massless vector-like pair supported on the corresponding matter
curve we need to be able to go beyond the chiral index and compute the individual numbers of left-
and right-chiral fermions for the $G_4$-flux \eqref{eq:fluxf11}. Unfortunately, these techniques are not
available as of now, see however \cite{Bies:2014sra} for promising recent advancements in this
direction. Thus, we work in the following under the assumption that the desired vector-like pair is indeed
part of the massless spectrum. Then it would be possible to induce
the following bilinear coupling
\begin{align} \label{eq:Wmu}
\mathcal{W} \subset \mu H_u H_d + \beta^i H_u L_i \; .
\end{align}
These two terms could be generated by tuning the complex structure of our model
to a model with enhanced (non-Abelian or Abelian) gauge symmetry and a SM-singlet $\mathbf{1}$,
that admits Yukawa couplings with $H_u$, $H_d$ and $L_i$, respectively. Then if $\mathbf{1}$
acquires a VEV, which breaks the enhanced gauge symmetry, the superpotential \eqref{eq:Wmu} could
be generated. While the $\mu$-term has to be very small in order to be consistent with electroweak
symmetry breaking, the $\beta^i$ terms are lepton violating and hence they must be adequately
suppressed.
We note that both these coefficients are moduli dependent functions, that cannot be
computed by known techniques. However, we expect that in a sufficiently generic geometry the moduli
of  $X_{F_{11}}$ allow for
appropriate tunings providing  a phenomenologically viable  scenario.   At this point, we must remark
that the geometry of  $X_{F_{11}}$ offers no obvious way by which we could assign a quantum
number to forbid the $\mu$-term or the $\beta^i$ terms.

Regarding the trilinear couplings we note that it was shown in \cite{Klevers:2014bqa} that all gauge
invariant trilinear couplings are realized geometrically, see Table \ref{tab:poly11_yukawa}.
\begin{table}[H]
\begin{center}
\renewcommand{\arraystretch}{1.3}
\begin{tabular}{|c|c|}\hline
Yukawa & Locus \\ \hline
$(\three,\two)_{1/6}	\cdot (\overline{\three},\one)_{-2/3}	\cdot \overline{(\one,\two)_{-1/2}}$ & $s_3=s_5=s_9=0$ \\ \hline
$(\three,\two)_{1/6}	\cdot (\overline{\three}, \one)_{1/3}	\cdot (\one,\two)_{-1/2}$ & $s_3=s_9=0=s_1s_6-s_2s_5$ \\ \hline
$(\overline{\three},\one)_{-2/3}	\overline{\cdot (\overline{\three}, \one)_{1/3}}	\cdot (\one,\one)_{1}$ & $s_1=s_5=s_9=0$ \\ \hline
$(\three,\two)_{1/6} \cdot (\three,\two)_{1/6} \cdot \overline{(\overline{\three},\one)_{1/3}}$ & $s_3=s_9=s_6=0$\\ \hline
$(\one,\two)_{-1/2} \cdot (\one,\two)_{-1/2} \cdot (\one,\one)_{1}$ & $s_1=s_5=s_3=0$\\ \hline
$(\overline{\three},\one)_{1/3} \cdot (\overline{\three},\one)_{1/3} \cdot (\overline{\three},\one)_{-2/3}$ & $s_5=s_6=s_9$\\ \hline
\end{tabular}
\caption{\label{tab:poly11_yukawa}Codimension three loci and corresponding Yukawa couplings for $X_{F_{11}}$.}
\end{center}
\end{table}
\noindent Thus, all MSSM Yukawas are geometrically allowed, giving rise to the superpotential terms
\begin{align} \label{eq:Yukawas}
\mathcal{W} \subset Y^{u}_{i,j} Q_i \overline{u}_j H_u +  Y^{d}_{i,j} Q_i \overline{d}_j H_d + Y^{L}_{i,j} \overline{e}_i \overline{L}_j H_d \, .
\end{align}
Since all three copies of each SM field live on the same matter curve, it is expected that the hierarchies in the
Yukawas are generated in a similar fashion as in most $\SU5$ F-theory GUTs, where the Yukawa matrix
has rank one, so that the geometrical coupling gives the mass for the heavy generation while the lighter
ones pick their masses from instanton contributions \cite{Font:2008id,Heckman:2008qa,Font:2012wq}

Note also that since we cannot distinguish between $H_d$ and $L_i$, the following dimension four proton decay operators are also geometrically allowed
\begin{align}\label{pdecay}
\mathcal{W} \subset & \lambda^{(0)}_{i,j,k}  Q_i \overline{d}_j L_k +\lambda^{(1)}_{i,j,k} \overline{e}_i L_j L_k + \lambda^{(2)}_{i,j,k} \overline{u}_i \overline{d}_j \overline{d}_k \, .
\end{align}
Here the coupling $\lambda^{(2)}$ can only be suppressed by appropriate tunings of moduli. However, this is not possible for $\lambda^{(0)}$, $\lambda^{(1)}$ as their geometric origin is the same as that
of the Yukawa couplings \eqref{eq:Yukawas}. It then seems very challenging to suppress them to orders of $\lambda \leq  10^{-10}$ \cite{Barbier:2004ez}, while keeping the Yukawa couplings at values of about 0.1.

\section{Pati-Salam Model: $G_{F_{13}}=(\text{SU(4)}\times\text{SU(2)}^2)/\mathbb{Z}_2$}\label{sec:pati}

The fibration $X_{F_{13}}$ has been shown to exhibit the gauge symmetry and matter representation 
of Pati-Salam (PS) model \cite{Klevers:2014bqa}. In Section \ref{sec:geomF13} we review the geometrical 
properties of $X_{F_{13}}$. In Section \ref{sec:G4F13} we explicitly construct  $G_4$-flux for the  
base $B=\mathbb{P}^3$. There we also compute the homology class of all matter surfaces and  the 
4D chiralities for all matter representations.  We also determine the minimal number of generations 
which allow for D3-brane tadpole cancelation and integral $G_4$-flux. Finally, the phenomenology of 
F-theoretic PS-models with three generations and their Higgsing down to the MSSM are described in Section \ref{sec:phenoF13}.

Readers directly interested in the 4D chiralities and phenomenological aspects of the models 
can directly jump to \eqref{eq:124} and the following discussions.

\subsection{The Geometry of Gauge Group and Matter Representations}\label{sec:geomF13}
\begin{figure}[H]
\centering
\begin{minipage}{.56\textwidth}
  \centering
  \includegraphics[scale=.35]{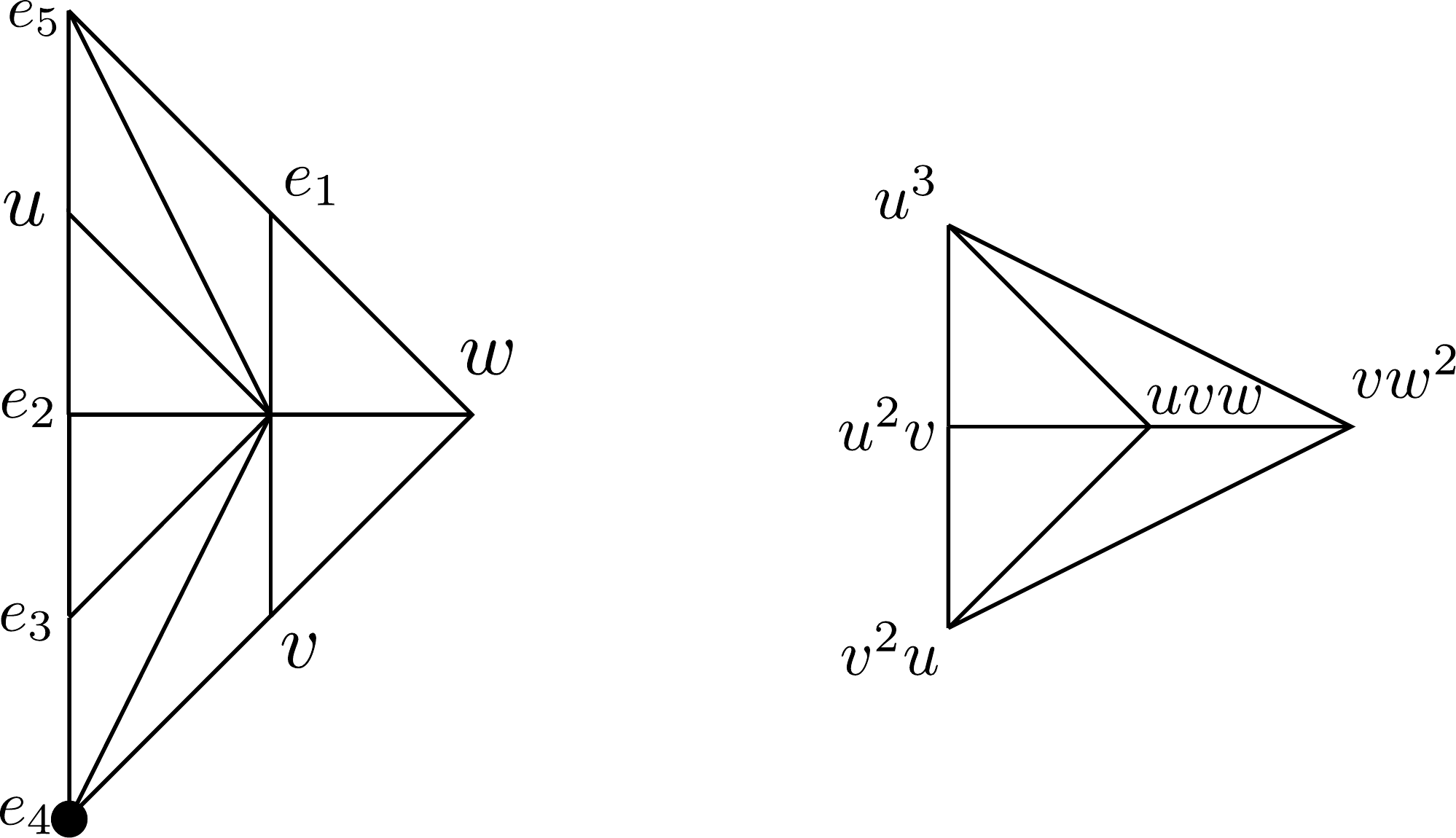}
\end{minipage}%
\begin{minipage}{.44\textwidth}
{\footnotesize
  \begin{tabular}{|c|c|}\hline
Section & Line Bundle\\ \hline
$u$ & $\mathcal{O}(H-E_1-E_2-E_5+\cS_9+[K_B])$ \\ \hline
$v$ & $\mathcal{O}(H-E_2-E_3-E_4+\cS_9-\cS_7)$\\ \hline
$w$ & $\mathcal{O}(H-E_1)$\\ \hline
$e_1$ & $\mathcal{O}(E_1-E_5)$\\ \hline
$e_2$ & $\mathcal{O}(E_2-E_3)$\\ \hline
$e_3$ &  $\mathcal{O}(E_3-E_4)$\\ \hline
$e_4$ &  $\mathcal{O}(E_4)$\\ \hline
$e_5$ &  $\mathcal{O}(E_5)$\\ \hline
\end{tabular}}
\end{minipage}
\caption{\label{fig:poly13_toric}The toric diagram of polyhedron $F_{13}$ and its dual. The zero section is indicated by the dot. In the accompanying table we indicate the divisor classes of the fiber coordinates.}
\end{figure}

The elliptic fiber which is yields F-theory models that naturally give rise to the gauge group and charge 
pattern needed for models of Pati-Salam (PS) unification is given by the following CY-hypersurface
\begin{align}
\begin{split} \label{eq:pF13}
p_{F_{13}}&= s_1 e_1^2 e_2^2e_3 e_5^4 u^3 + s_2 e_1 e_2^2 e_3^2 e_4^2 e_5^2 u^2 v + s_3 e_2^2 e_3^3 e_4^4 u v^2 + s_6 e_1 e_2 e_3 e_4 e_5 u v w 
+s_9 e_1 v w^2\, ,
\end{split}
\end{align}
defined in the toric ambient space $\mathbb{P}_{F_{13}}$. The toric diagram of the ambient space as well as the divisor classes of the fiber coordinates are summarized in Figure \ref{fig:poly13_toric},
where as before $H$ is the hyperplane on $\mathbb{P}^2$ and $E_i$, $i=1,\ldots,5$,  denote the exceptional divisors.

The elliptically fibered Calabi-Yau fourfold $X_{F_{13}}$ is constructed  by promoting the coefficients $s_i$ to sections of the line bundles over the base $B$ given in \eqref{eq:cubicsections}.  The elliptic 
fibration of $X_{F_{11}}$ is equipped with a zero section given by
\begin{align}
\hat{s}_0=X_{F_{13}}\cap\{e_4=0\}&:\quad [1:s_1:1:1:1:-s_9:0:1]\, .
\end{align}
Furthermore, the elliptic fibration admits a section of order two, giving rise to $\mathbb{Z}_2$ 
Mordell-Weil group \cite{Mayrhofer:2014opa,Klevers:2014bqa}.
In addition, one can see by computing the discriminant that over the locus 
$\mathcal{S}_{\SU4}=\{s_9=0\}$ the fiber degenerates to an $I_4$-fiber. The corresponding Cartan 
divisors of the resulting $\SU4$ gauge symmetry are given by
\begin{align}
D^{\SU4}_{1}=[s_9]-[u]-[e_2]-[e_3]\, , \quad D^{\SU4}_{2}=[u] \, ,\quad D^{\SU4}_{3}=[e_2] \, .
\end{align}
Similarly, at the loci $\mathcal{S}_{\SU2_1}=\{s_1=0\}$ and $\mathcal{S}_{\SU2_2}=\{s_3=0\}$ we obtain $I_2$-fibers. The resulting two $\SU2$ factors have the following associated Cartan divisors:
\begin{align}
D^{\SU2_1}_{1}=[s_1]-[v] \,,\quad
D^{\SU2_2}_{1}=[e_1] \, .
\end{align}

The codimension two loci where the singularities of the fibration enhance, corresponding to the 
presence of matter fields, are given in Table \ref{tab:poly13_matter}. We readily observe that the 
codimension two loci of $X_{F_{13}}$ support the matter representations characteristic for the 
Pati-Salam model.
\begin{table}[H]
\begin{center}
\renewcommand{\arraystretch}{1.2}
\begin{tabular}{|c|c|}\hline
Representation  & Locus \\ \hline

$(\one,\two, \two)$ & $V(I_{(1)}):=  \{ s_1=s_3=0 \}$  \\ \hline

$(\mathbf{4},\two, \one)$ & $V(I_{(2)}):= \{ s_1=s_9=0 \}$  \\ \hline

$(\mathbf{4},\one, \two)$ & $V(I_{(3)}):= \{ s_3=s_9=0 \}$  \\ \hline

$(\mathbf{6},\one, \one)$ & $V(I_{(4)}):= \{ s_6=s_9=0 \}$  \\ \hline
\end{tabular}
\caption{\label{tab:poly13_matter}Charged matter representations under $(\SU4\times\SU2^2)/\mathbb{Z}_2$ and their associated codimension two loci on $X_{F_{13}}$.}
\end{center}
\end{table}
Similar as in section \ref{sec:geomF11} we provide base independent expressions for the second 
Chern class as well as the Euler number of $X_{F_{13}}$, which are needed for $G_4$-flux quantization
and D3-brane tadpoles. We obtain
\begin{align}
  \begin{split}
  c_2(X_{F_{13}})&= -c_1^2 + c_2 - 6 c_1 E_1 - 6 E_1^2 - 2 c_1 E_3 - 2 E_4^2 + 4 c_1 H + c_1 \cS_7 + 9 E_1 \cS_7 + E_2 \cS_7 \\
    &\phantom{=} + E_3 \cS_7 - E_4 \cS_7 - E_5 \cS_7 - 3 H \cS_7 - 4 E_1 \cS_9 - 2 E_2 \cS_9 + 2 H \cS_9 - 2 \cS_7 \cS_9 + 2 \cS_9^2\; ,
  \end{split}\\
  \chi(X_{F_{13}})&= 12 (6 c_1^3+c_1 c_2-4 c_1^2 \cS_7+2 c_1 \cS_7^2-6 c_1^2 \cS_9+2 c_1 \cS_7 \cS_9-\cS_7^2 \cS_9+2 c_1 \cS_9^2) \label{eq:eulerf13}\; ,
\end{align}
where, as before, $c_1$ and $c_2$ denote the first and second Chern class of the base $B$, 
respectively, and the divisors $\mathcal{S}_7$ and $\mathcal{S}_9$ are introduced in 
\eqref{eq:cubicsections}. 

\begin{figure}[H]
\center
\includegraphics[scale=0.8]{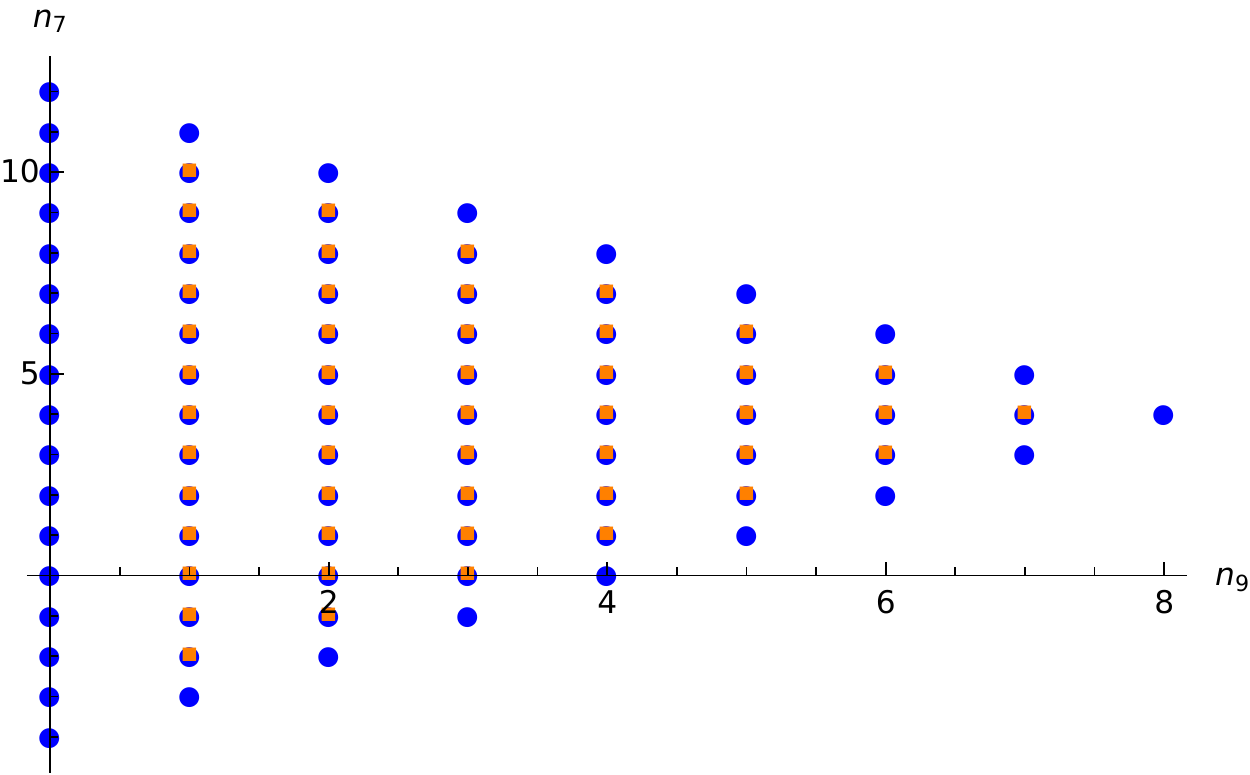}
\caption{\label{fig:f13allowed}Allowed region of $(n_7,n_9)$ for the CY-fourfold $X_{F_{13}}$ 
with $B=\mathbb{P}^3$. For points in orange, all representations of the PS-model
are present and a $G_4$-flux admitting  $b$  families exists.}
\end{figure}

Again, we fix the base of the fibration to be $B=\mathbb{P}^3$ for the remainder of this section. 
We expand the divisor $\mathcal{S}_7$, $\mathcal{S}_9$ and the anti-canonical class $[K_{B}^{-1}]$
w.r.t.~$H_B$ as in \ref{eq:baseP3}. Demanding effectiveness of all  sections in 
\eqref{eq:cubicsections} entering the CY-constraint \eqref{eq:pF13}, we find the allowed values for the 
pair $(n_7,n_9)$ depicted in Figure \ref{fig:f13allowed}.
\subsection{$G_4$-Flux and Chiral Generations}\label{sec:G4F13}
For the base $B=\mathbb{P}^3$, the SR-ideal  of the toric ambient space
of the fourfold $X_{F_{13}}$ is given by
\begin{align}
\begin{split}\label{eq:SRF13}
SR_{F_{13}}&=\{ u e_1, uw, uv, u e_4, u e_3, e_5 w, e_5 v, e_5 e_4, e_5 e_3, e_5 e_2, e_1 v, e_1 e_4, e_1 e_3, e_1 e_2, \\
&\phantom{=...} w e_4, w e_3, we_2, ve_3, ve_2, e_4 e_2, x_0x_1x_2x_3\} \, ,
\end{split}
\end{align}
where, again, $[u,v,w]$ and the $e_i$, $i=1,\ldots,5$, are the projective coordinates on the fiber and 
$x_j$ ($j=0,1,2,3$) are the homogeneous coordinates on the $\mathbb{P}^3$ base.  As a basis of 
$H^{1,1}(X_{F_{13}})$  we choose the hyperplane $H_B$, the zero section and the five Cartan 
divisors,
\begin{align} \label{eq:H11F13}
H^{(1,1)}(X_{F_{13}})&= \langle H_B, S_0, D^{\SU2_1}_{1}, D^{\SU2_2}_{1}, D^{\SU4}_{1}, D^{\SU4}_{2}, D^{\SU4}_{3} \rangle \; ,
\end{align}
where, as before,  the divisors and their dual $(1,1)$-forms are denoted by the same symbol. Here, 
$S_0$ is the class of $\hat{s}_0$.

For the computation of the vertical cohomology ring of $X_{F_{13}}$ we use the SR-ideal \eqref{eq:SRF13}, the basis \eqref{eq:H11F13} as well as the following intersections,
\begin{align}
H_B^3\cdot S_0^2=-1\;,\quad H_B^3\cdot \big(D^{\SU2_2}_{1}\big)^2=-2\;,\quad H_B^3\cdot \big(D^{\SU4}_{3}\big)^2=-2\;,
\end{align}
which follow from intersections in   $\mathbb{P}_{F_{13}}$. We readily obtain the quartic intersections 
of $X_{F_{13}}$ and the basis of $H^{(3,3)}(X_{F_{13}})$ is canonically determined by
\eqref{eq:H11F13}.
Considering all products of two divisors in $H^{(1,1)}(X_{F_{13}})$, we obtain generators for
$H^{(2,2)}_V(X_{F_{13}})$. The rank of their inner product matrix reveals that 
\begin{align}
{\rm dim} H^{(2,2)}_V(X_{F_{13}}) &=8 \; .
\end{align}
We choose the following eight-dimensional basis for $H^{(2,2)}_V(X_{F_{13}})$:
\begin{align}
H^{(2,2)}_V(X_{F_{13}})&= \langle H_B^2, H_B S_0, H_B D^{\SU2_1}_{1}, H_B D^{\SU2_2}_{1}, H_B D^{\SU4}_{1}, H_B D^{\SU4}_{2}, H_B D^{\SU4}_{3} , S_0^2 \rangle \; .
\end{align}

Next, we make an ansatz for the $G_4$-flux in terms of this basis. Following the description in Section 
\ref{sec:review} we impose the condition \eqref{eq:csfluxconstraints}. We find seven conditions on the 
$G_4$-flux which leaves us with the following one parameter flux:
\begin{align}\label{eq:fluxf13}
\begin{split}
G_4 &=-a_8\left[S_0^2+H_B\cdot\left(4 S_0 + (12 n_9 -n_7 n_9-n_9^2)H_{B} -\tfrac12 n_9D^{\SU2_1}_{1} \right.\right.\\
&\phantom{=}\,\,\left.\left.+ \tfrac14( n_7 +n_9-12)(3 D^{\SU4}_{1}+2 D^{\SU4}_{2}+ D^{\SU4}_{3}) \right)\right]\;.
\end{split}
\end{align}
Here $a_8$ is a free discrete parameter, whose quantization is determined by $G_4$-flux quantization.
As in section \ref{sec:standardmodel} we will address $G_4$-flux quantization indirectly in
dependence on the number of chiral families by ensuring an integral and positive number
$n_{\rm D3}$ of D3-branes and quantization of  3D CS-terms \eqref{eq:thetaM}. We present the 
findings of this analysis below.

As a next step, we need the homology classes of all matter surfaces for the representations in Table 
\ref{tab:poly13_matter}. Two of the matter surfaces are given as complete intersections in the toric 
ambient space $X_{F_{13}}$. Their homology classes are given by
\begin{align}
\begin{split}
\mathcal{C}^w_{(\four,\two,\one)}&= \cS_9\cdot (3 [K_B^{-1}] - \cS_7 - \cS_9) \cdot(H + [K_B^{-1}])\;,\quad 
\mathcal{C}^w_{(\four,\one,\two)}= \cS_9 ([K_B^{-1}] + \cS_7 - \cS_9) E_5\;,
\end{split}
\end{align}
where we chose those nodes which are not intersected by the zero section.
These matter surfaces lead to the following chiralities
\begin{align}
\chi_{(\four,\two,\one)}=-\chi_{(\four,\one,\two)}=\frac{1}{4} n_9 \left(4+n_7-n_9 \right)\left(n_7+n_9 -12 \right)a_8\; .\label{eq:124}
\end{align}

We note that the representations $(\one,\two,\two)$ and $(\six,\one,\one)$ (see Table 
\ref{tab:poly13_matter}) are real representations. Therefore, their chiralities \eqref{eq:g4chiralities} 
are zero by definition. This is manifested in the geometry of $X_{F_{13}}$ by the fact that the
fiber at the corresponding codimension two loci is non-split \cite{Klevers:2014bqa}, so that precisely
the nodes carrying the weights of the respective representation are interchanged by codimension
three monodromies. This implies that the integral of the $G_4$-flux over the matter surfaces 
associated to one note is equal to the integral over the matter surface associated to the other node.
However, as the sum of the weights of the two nodes has to equal a root of the gauge group, the
two integrals add up to zero, see \eqref{eq:csfluxconstraints}, and we get zero chirality. 
It is important to stress that this does not necessarily mean that there are no massless fields 
transforming under real representations, in F-theory. Indeed, there can be vector-like
fields in the theory, which are counted by the number $n(\mathbf{R})$ of individual
Weyl fermions.  For the above reasons, however, the multiplicity of  fermions in real representations is 
not accessible by the methods described in this work. It would be interesting to develop methods based
on those introduced in \cite{Bies:2014sra}  to compute the number $n(\mathbf{R})$ for real representations 
explicitly.

We note that according to \eqref{eq:124}, the
chiralities of the two complex representations $(\four,\two,\one)$ and $(\four,\one,\two)$
are equal up to a sign, which guarantees the cancelation of the cubic \SU4 
anomaly.\footnote{Note that the cubic \SU4 anomaly, together with the Witten anomaly of the \SU2 
factors are the only anomalies to check in this model. One can also see that the cancelation of the \SU4 anomaly guarantees that the Witten anomaly cancels as well.} Thus,  the number $b$ of chiral 
generations coincides with the chirality of one of these two representations, i.e.~we set $b=
\chi_{(\four,\two,\one)}$. This allows us to express the parameter $a_8$ in the $G_4$-flux in 
\eqref{eq:fluxf13} as
\begin{align}\label{eq:familyparameterf13}
a_8 =  \frac{ 4 b}{n_9 \left(n_9-n_7-4\right)\left(12-n_7-n_9  \right)} \, .
\end{align}
We note again that those values for $(n_7,n_9)$, for which the denominator vanishes, are 
excluded by the requirement of non-vanishing chiralities \eqref{eq:124}. Thus, we are restricted to the  
interior  (orange dots) of the allowed region for the fibration over $\mathbb{P}^3$ in  Figure \ref{fig:f13allowed}. 

As in Section \ref{sec:anomalien},  we parametrize the $G_4$-flux in terms of the integral number $b$ 
in \eqref{eq:familyparameterf13} of families. 
In this parametrization we evaluate the D3-brane tadpole \eqref{eq:tadpole} and require integral, 
positive $n_{\rm D3}$ and integral CS-terms \eqref{eq:thetaM}. Without further horizontal $G_4$-flux, 
this yields the lower bounds on the number $b$ in dependence of $(n_7,n_9)$ shown in Table 
\ref{tab:families_f13}, where also the corresponding numbers $n_{\rm D3}$ of D3-branes are 
displayed. We find that three families are possible at two values for $n_7$ and $n_9$. Again, we find 
that in the context of this simple analysis three is the minimal number $b$ of families 
compatible with the D3-brane tadpole \eqref{eq:tadpole}. We also note that Table 
\ref{tab:families_f13} has inherited the symmetries of the polyhedron $F_{14}$.
\begin{table}
\begin{center}
\renewcommand{\arraystretch}{1.4}
{\footnotesize
\begin{tabular}{c|ccccccc}
 {\large $_{n_7} \backslash ^{n_9}$} & 1 & 2 & 3 & 4 & 5 & 6 & 7 \\ \hline
  10 & $( 13;204 )$ &  &  &  &  &  &  \\
 9 & {\huge -} & $( 11;140 )$ &  &  &  &  &  \\
  8 & $( 33;94 )$ & $( 10;119 )$ & $( 9;90 )$ &  &  &  &  \\
 7 & {\huge -} & $( 9;100 )$ & $( 6;77 )$ & $( 14;48 )$ &  &  &  \\
 6 & $( 15;108 )$ & $( 8;86 )$ & $( 21;52 )$ & $( 12;46 )$ & $( 5;44 )$ &  &  \\
 5 & $( 6;106 )$ & $( 35;44 )$ & {\huge -} & $( 30;16 )$ & {\huge -} & {\color{red}$( 3;44 )$} &  \\
 4 & $( 7;102 )$ & $( 6;75 )$ & $( 15;50 )$ & $( 8;42 )$ & $( 15;30 )$ & $( 6;41 )$ & $( 7;42 )$ \\
  3 & $( 6;106 )$ & $( 35;44 )$ & {\huge -} & $( 30;16 )$ & {\huge -} & {\color{red}$( 3;44 )$} & \\
 2 & $( 15;108 )$ & $( 8;86 )$ & $( 21;52 )$ & $( 12;46 )$ & $( 5;44 )$ &  &  \\
 1 & {\huge -} & $( 9;100 )$ & $( 6;77 )$ & $( 14;48 )$ &  &  &  \\
 0 & $( 33;94 )$ & $( 10;119 )$ & $( 9;90 )$ &  &  &  &  \\
 -1 & {\huge -} & $( 11;140 )$ &  &  &  &  &  \\
  -2 & $( 13;204 )$ &  &  &  &  &  & \\
\end{tabular}}
\caption{\label{tab:families_f13} The entries $(b;n_{\rm D3})$ show the minimal number of families 
$b$ for which the number of D3 branes $n_{D3}$ is integral and positive for integral 3D CS terms. At 
the points marked with "-" the number of D3 branes is negative
for all positive integral values of $b$.}
\end{center}
\end{table}

We conclude by noting that we double-check the chiralities in \eqref{eq:124} using the matching
\eqref{eq:M=F} of CS-terms.  The real representations in Table \ref{tab:poly13_matter} do not
contribute to the loop-corrections in \eqref{eq:thetaF}. We obtain the non-vanishing CS-terms
\beq
	\Theta^\mathrm{F}_{i,i}=-2 \chi_{\four,\one, \two}\,\,\,\, (i=1,\ldots,5)\,,\,\qquad  \Theta^\mathrm{F}_{i=3,j=4} =\Theta^\mathrm{F}_{i=4,j=5}=-\chi_{\four,\one, \two}\,,
\eeq
where the indices label the five Cartan divisor 
$\{D^{\SU2_1}_{1}, D^{\SU2_2}_{1}, D^{\SU4}_{1}, D^{\SU4}_{2}, D^{\SU4}_{3}\}_i$. Equating this with the CS-terms \eqref{eq:thetaM} on the M-theory 
side we readily reproduces \eqref{eq:124}.

\subsection{Phenomenological Discussion}\label{sec:phenoF13}

In Table \ref{tab:families_f13} we find only two models with $(n_7 , n_9)=(5,6),\, (3,6)$ that allow for three chiral Pati-Salam families. As in the standard model we see that three is the minimum allowed number of generations. 
These two models are equivalent under the reflection along the $(4,n_9)$ line, which reflects the invariance of the theory under exchange of the two SU(2) gauge groups.

The Higgs transition from $X_{F_{13}}$ to $X_{F_{11}}$ has been considered  for the six dimensional case 
\cite{Klevers:2014bqa}. However, some of the observations made there immediately carry over to four dimensions. Similar to 
the 6D case, the transition happens due to a toric blow-down in the ambient space of the elliptic fiber of $X_{F_{13}}$. In this 
case we see that blowing down either $e_4$ or $e_5$ in Figure \,\ref{fig:poly13_toric} leads to the toric diagram of $F_{11}$ in 
Figure \ref{fig:poly11_toric}. These two transitions are equivalent up to redefinitions of the coordinates on the fiber, and for this 
reason we focus only on the blow down of $e_4$ which leads to $F_{11}$ in its canonical form. 

There are some subtleties that have to be discussed before proceeding with the detailed discussion of the Higgsings.
First, we note that the blow-down process requires a restriction of the allowed region in Figure \ref{fig:f13allowed}  for 
$X_{F_{13}}$ because the section $s_5$ has to be effective, which is present in the CY-equation \eqref{eq:pF11} for   
$X_{F_{11}}$, but not in the one \eqref{eq:pF13} for  $X_{F_{13}}$. This reduction amounts to excluding only two models 
below the $(8,n_9)$ line in Table \ref{tab:families_f16}. To understand from an effective field theory point view, why the 
Higgsing is not possible for these two models is elusive. We 
recall that in the six dimensional case, the exclusion of 
certain points in the allowed region for $(n_7,n_9)$ is explained from the field theory point of view by the fact that these 
models  lack a sufficient number of Higgses for carrying out a D-flat Higgsing. 
In four dimensions one expects the same to occur. However, here the Higgses are pairs of vector-like fields and due to our 
lack of control over this sector of the theory, we can not verify this statement at this point. 

Second, we also note that while in $X_{F_{11}}$ certain points in Table \ref{tab:families_f11} do not permit a family 
structure, in $X_{F_{13}}$ some of these points do allow for a certain number of families. Since the net chirality is preserved by 
Higgsings, it remains the question of why it is not possible to have the Higgsed models from $X_{F_{13}}$ as consistent tadpole 
canceling solutions\footnote{A logically possible explanation is again the absence of enough Higgs fields for performing the 
transition to begin with, which can not be tested with the tools at hand.} 
in $X_{F_{11}}$. It is expected that this seeming contradiction can be resolved by the inclusion of horizontal $G_4$-flux. For 
example, it has been argued in \cite{Intriligator:2012ue,Mayrhofer:2014opa}, at least in simple situations, 
that horizontal $G_4$-flux exists that compensates for the change of the Euler number of a CY-fourfold 
in an extremal transition. Adding this $G_4$-flux to F-theory on $X_{F_{11}}$, its D3-brane tadpole \eqref{eq:tadpole}
becomes effectively identical to that on $X_{F_{13}}$ and we expect that our search strategy for three family models on both 
$X_{F_{11}}$ and $X_{F_{13}}$ will be consistent with the Higgs effect relating their effective actions.

In the following, we consider the special point $(n_7,n_9)=(5,6)$ which has three as the smallest number of families both in 
$X_{F_{11}}$ and $X_{F_{13}}$. Therefore, the toric Higgsing is possible without the necessity of adding further 
$G_4$-flux.\footnote{Note that the point $(n_7,n_9)=(2,5)$ allows for three families in 
$X_{F_{11}}$, while in $X_{F_{13}}$ the smallest number is five. Hence, based on these simple arguments the Higgsing is not 
possible without discussing horizontal $G_4$-flux.}  We discuss the field theoretical Higgsing in more detail and make 
some remarks about the phenomenology of this  three-family model.

On the field theory side the Higgsing is triggered by a VEV in the $(\four,\one,\two )$ representation at the locus $s_3 = s_9 = 0$. A supersymmetric Higgsing requires at least one vector-like pair of fields in the representation $(\four,\one,\two )$ in addition to the three chiral families. However, as mentioned already, we can not determine the massless vector-like spectrum with the techniques presented in section \ref{sec:review}. Therefore, we  work under the assumption that this vector-like pair of Higgs fields is indeed part of the spectrum. 
\begin{table}[H]
\begin{center}
\renewcommand{\arraystretch}{1.2}
\begin{tabular}{|c|ccl@{ }l@{ }l@{ }l|}\hline
Name & Representation &  & \multicolumn{4}{c|}{SM decomposition} \\ \hline
PS Higgs & $ 1\times(\overline{\four},\one, \two )_{H^1}$ & $\rightarrow$ & $ \bar{d}_{H}: (\overline{\three},\one)_{\frac13}$, & $\bar{u}_{H}: (\overline{\three},\one)_{-\frac23}$, & $\bar{e}_{H}: (\one,\one)_{1}$, & $\bar{\nu}_{H}: (\one,\one)_{0}$   \\
PS Higgs & $1\times (\four,\one, \two )_{H^2}$ &  $\rightarrow$ & $d_{H}: (\three,\one)_{-\frac13}$, & $u_{H}: (\three,\one)_{\frac23}$, & $e_{H}: (\one,\one)_{-1}$, & $\nu_{H}: (\one,\one)_{0}$ \\
Exotic & $1\times (\six,\one,\one)$ & $ \rightarrow $& $\bar{D}: (\overline{\three},\one)_{\frac13}$, & $D: (\three,\one)_{-\frac13}$ & &\\ \hline
SM Higgs & $1\times (\one,\two,\two)_{H}$ & $\rightarrow$ & $H_u: (\one,\two)_{1/2}$, & $H_d: (\one, \two)_{-\frac12}$ & &\\
SM Matter & $3 \times (\four, \two , \one)_{M}$ & $\rightarrow$ &  $Q_i: (\three,\two)_{\frac16}$, & $L_i: (\one,\two)_{-\frac12} $ & &\\
SM Matter & $3 \times (\overline{\four}, \one , \two)_{M}$ & $\rightarrow$ & $\overline{d}: (\overline{\three},\one)_{\frac13}$, & $\overline{u}: (\overline{\three},\one)_{-\frac23}$, & $\overline{e}: (\one,\one)_{1}$, & $\bar{\nu}: (\one,\one)_{0}$ \\
 \hline
\end{tabular}
 \caption{\label{tab:PSdecomposition}	The Pati-Salam matter content and its decomposition into standard model fields. The MSSM spectrum originates purely from the PS representations of the last three rows, which also provide candidates for right handed neutrinos $\nu$. The representation in the first three rows are the Pati-Salam Higgses and the sextet needed to decouple exotic triplets.}
\end{center}
\end{table}
In Table \ref{tab:PSdecomposition} we summarize the desired spectrum of the PS-model and its decomposition in terms of 
representations under the SM gauge group. In addition to the three chiral pairs 
$(\overline{\four},\one, \two )_{M}$, $(\four,\two,\one)_{M}$, we require the presence of a light vector-like pair 
$(\overline{\four},\one, \two )_{H^1}$ and $(\four,\one, \two )_{H^2}$ to serve as the PS-Higgses, whose neutral components 
develop VEVs $\langle H^1\rangle=\langle H^2\rangle$. In this type of breaking the hypercharge generator of the SM
corresponds to the combination
\begin{align}
Q_Y = {\textstyle \frac{2}{\sqrt{6}}}T_4^{15} - T^3_2 \, 
\end{align}
of the broken Cartans $T_4^{15}= {\textstyle \frac{1}{2\sqrt{6}}}\text{diag}(1,1,1,-3)$ in \SU4 and $T^3_2=\frac{\sigma^3}{2}$ in \SU2. Note also that in addition to these fields, we also expect a sextet and a bidoublet to be part of the massless spectrum\footnote{We recall again that the presence of real representations and  of a single vector-like pair of 
PS-Higgses is just introduced as an optimistic possibility in our discussion. It is to be seen if the hypersurface fibration over $\mathbb{P}^3$ actually allows for these desired fields.}. From the bidoublet $(\one,\two,\two)_H$ we get the SM-Higgses and from the $(\six,\one,\one)$ we get a pair of color triplets which serve to decouple some otherwise massless fields arising from the PS-Higgs multiplet \cite{Antoniadis:1988cm}.

\begin{table}[H]
\begin{center}
\renewcommand{\arraystretch}{1.3}
\begin{tabular}{|c|c|}\hline
Yukawa & Locus \\ \hline
$(\one,\two,\two) \cdot \overline{(\mathbf{4},\two,\one)} \cdot (\mathbf{4},\one,\two)$ & $s_1=s_3=s_9=0$ \\ \hline
$\overline{(\six,\one,\one)} \cdot (\mathbf{4},\two,\one) \cdot (\mathbf{4},\two,\one)$ & $s_1=s_6=s_9=0$ \\ \hline
$\overline{(\six,\one,\one)} \cdot (\mathbf{4},\one,\two) \cdot (\mathbf{4},\one,\two)$ & $s_3=s_6=s_9=0$ \\ \hline
\end{tabular}
\caption{\label{tab:poly13_yukawa}Codimension three loci and corresponding Yukawa couplings for $X_{F_{13}}$.}
\end{center}
\end{table}

Regarding the decomposition of the Higgs fields $H^1$ and $H^2$ in Table \ref{tab:PSdecomposition}, 
we note that the fields $\bar{u}_H$, $u_H$, $\bar{e}_H$, $e_H$ are removed from the massless spectrum as half of them 
become the longitudinal modes of the massive  bosons of the broken $\SU4\times\SU2$ and the other half become
massive Higgs bosons \cite{King:1997ia}. Therefore one only has to care about the lifting of the states $\bar{d}_H$ and $d_H$. 
For that purpose we use the exotic $(\six,\one,\one)$.  In Table \ref{tab:poly13_yukawa} we present the geometrically allowed 
Yukawa couplings in $X_{F_{13}}$. Writing the couplings involving the sextet and the Higgses and decomposing them into SM 
representations we find
\begin{align}
\langle H^1\rangle \bar{d}_H D+\langle H^2\rangle \bar{D} d_H \subset (\six,\one,\one)\cdot(\overline{\four},\one,\two)^2_{H^1}+(\six,\one,\one)\cdot(\four,\one,\two)^2_{H^2}\,.
\end{align}
With these couplings, the masses of all exotics can be pushed towards the grand unification scale $\Lambda_{\rm PS}$. 

For 
the real representations we generically expect the following field theoretical bilinears in the superpotential
\begin{align}
\label{eq:PSmassterms}
\mathcal{W} \supset& \, M_2 (\one,\two,\two)_H\cdot (\one,\two,\two)_H + M_6 (\six,\one,\one)\cdot(\six,\one,\one)\,
 \, .
\end{align}
Similar to the bilinear couplings \eqref{eq:Wmu} in the SM, these terms are expected to be generated by a VEV of a 
PS-singlet, that can be made visible by tuning the complex structure of $X_{F_{13}}$.\footnote{In fact, we can confirm, that 
there are codimension three components in the matter curves of the real representations of the PS-model, that could 
geometrically support the couplings \eqref{eq:PSmassterms}.}
Thus, the masses $M_2$ and $M_6$ are expected to be generically above the PS unification scale. However, after 
decomposing the bidoublets in terms of the SM gauge group, we see that the mass $M_2$ is related to the mass coefficient in 
front of the bilinear $H_u H_d$ in \eqref{eq:Wmu}. Therefore, some fine tuning is needed in order to guarantee that the $\mu$ 
term is small enough. Similarly, the coupling $M_6$ enforces a kind of see-saw mechanism for the triplets $d_H$ and $D$. 
The lowest mass eigenstate would have a mass of the order $\Lambda_{\rm PS}^2/M_6$. Hence, we have to guarantee that 
$M_6$ is not too far beyond the PS scale as otherwise the exotic masses will be pushed towards observable mass
scales. Therefore, we  have to rely on the possibility that certain points in the complex structure moduli space of $X_{F_{13}}$ 
allow for a configuration in which both $M_2$ and $M_6$ are sufficiently small.

The Yukawa couplings for the SM fields are also generated geometrically as one can see from Table \ref{tab:poly13_yukawa}. 
Indeed, we have the couplings
\begin{align}
\mathcal{W} & \supset (\one,\two,\two)_{H}\cdot (\four,\two,\one)_{M,i} \cdot (\overline{\four},\one,\two)_{M,j} \,.
\end{align}
Since both up-and down-type Yukawas arise from the same Yukawa point, the masses for up and down type quarks coincide 
at the PS-scale. The observed mass splitting is then due to the RG-running of the masses to the infrared. Again, since all 
matter fields of the same type arise from the same matter curve, the rank of the Yukawa matrix is expected to be one, with the 
lighter families picking their masses from instanton effects. 

We conclude with a final important remark about the Pati-Salam model. We note that Higgses and leptons arise from different 
representations of the PS-group, which is in contrast to GUT schemes such based on $\SO{10}$ or $\SU{5}$. This has the 
advantage that in PS-models, there is no Yukawa coupling which induces the dimension four proton decay operators 
\eqref{pdecay}, which are only generated below $\Lambda_{\rm PS}$ after integrating out the heavy triplets. Therefore, these 
couplings are suppressed by a factor $\Lambda_{\rm PS}/M_6$. The dimension five proton decay operators are generated in a 
similar fashion. To see this in more detail let us consider the couplings of the SM fields to the exotic sextet
\begin{align}
\begin{split}
(\six,\one,\one) (\overline{\four},\one,\two)_{M,i} (\overline{\four},\one,\two)_{M,j} & \supset \bar{D} \bar{d}_i \bar{u}_j+D \bar{d}_i \nu_j+D \bar{u}_i \bar{e}_j \,,\\
(\six,\one,\one) (\four,\two,\one)_{M,i} (\four,\two,\one)_{M,j} & \supset \bar{D}Q_iL_j+D Q_i Q_j\,.
\end{split}
\end{align}
Upon integration of the exotic states $D$, $\bar{D}$ we obtain the effective five operators, 
\begin{align}
Q_i Q_j Q_k L_m+ Q_i L_j \bar{u}_k \bar{e}_i+\bar{d}_i \bar{u}_j \bar{u}_k\bar{e}_m
\end{align}
all of which are suppressed by a factor $(1/M_6^2)$. 

\section{Trinification Model: $G_{F_{16}}=(\text{SU(3)}^3)/\mathbb{Z}_3$}\label{sec:trinif}
The last type of models we consider are F-theory compactifications on $X_{F_{16}}$, which exhibit the gauge group of 
Trinification as well as its characteristic bitriplet spectrum. The geometrical information relevant for the discussion of the matter 
and gauge content of the theory is provided in Section \ref{sec:geomF16}. The construction of the $G_4$-flux for this models 
and the induced 4D matter chiralities is discussed in Section \ref{sec:G4F16}. There we also determine the smallest allowed 
numbers of families for all of the allowed strata in moduli space. Finally,  in Section \ref{sec:phenoF16} we focus on a particular 
model which allows for three generation. We describe the Higgsings to the F-theoretic SM obtained from $X_{F_{11}}$, both 
on the geometry and field theory sides, and comment on the phenomenology of the model.

Readers directly interested in the 4D matter chiralities and the subsequent phenomenological discussion can
start reading at \eqref{chiF16}. 

\subsection{The Geometry of  Gauge Group and Matter Representations}\label{sec:geomF16}
\begin{figure}[H]
\centering
\begin{minipage}{.57\textwidth}
  \centering
  \includegraphics[scale=.4]{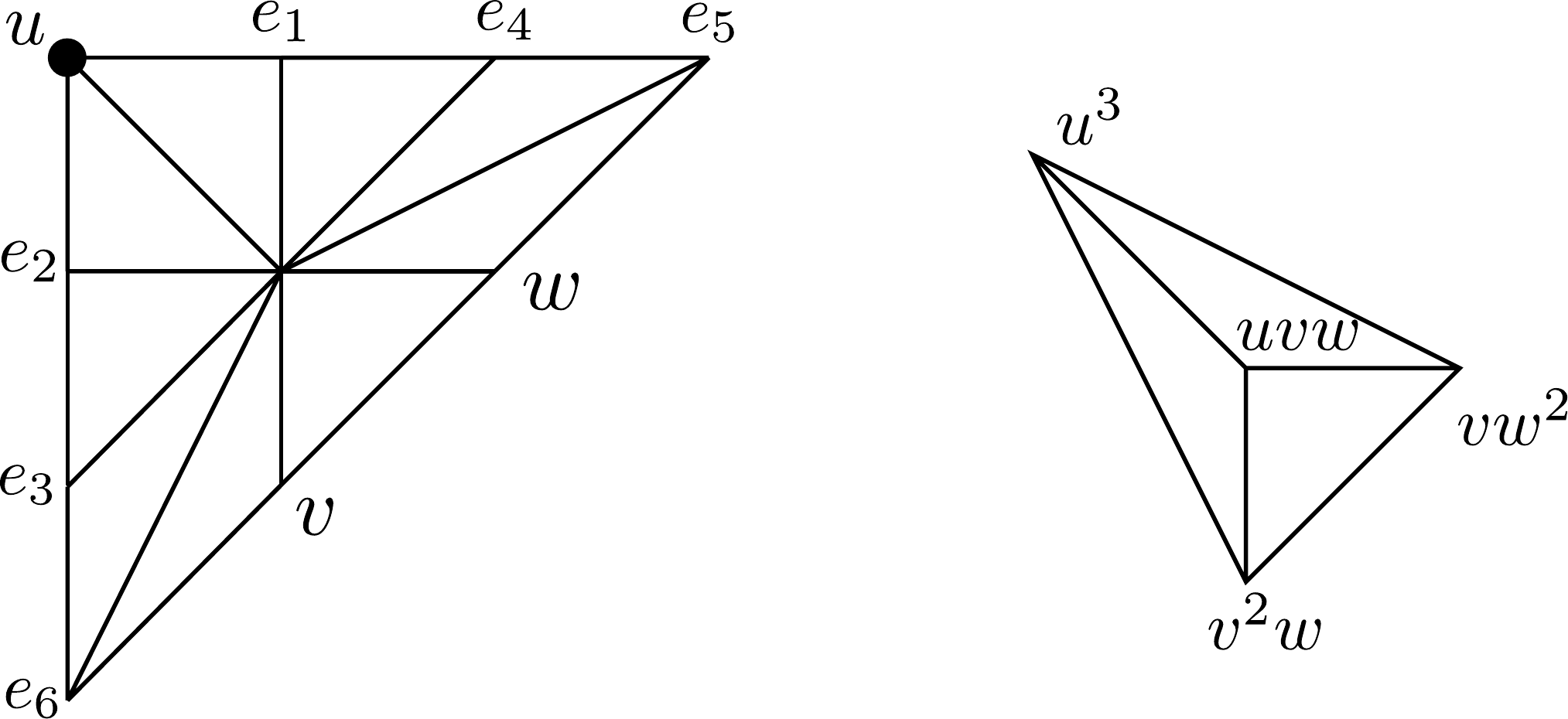}
\end{minipage}%
\begin{minipage}{.43\textwidth}
{\footnotesize
  \begin{tabular}{|c|c|}\hline
Section & Divisor class\\ \hline
$u$ & $\mathcal{O}(H-E_1-E_2+\cS_9+[K_B])$ \\ \hline
$v$ & $\mathcal{O}(H-E_2-E_3-E_6+\cS_9-\cS_7)$\\ \hline
$w$ & $\mathcal{O}(H-E_1-E_4-E_5)$\\ \hline
$e_1$ & $\mathcal{O}(E_1-E_4)$\\ \hline
$e_2$ & $\mathcal{O}(E_2-E_3)$\\ \hline
$e_3$ &  $\mathcal{O}(E_3-E_6)$\\ \hline
$e_4$ &  $\mathcal{O}(E_4-E_5)$\\ \hline
$e_5$ &  $\mathcal{O}(E_5)$\\ \hline
$e_6$ &  $\mathcal{O}(E_6)$\\ \hline
\end{tabular}}
\end{minipage}
\caption{\label{fig:poly16_toric}The toric diagram of polyhedron $F_{16}$ and its dual. The zero section is indicated by the dot. In the accompanying table we indicate the divisor classes of the fiber coordinates.}
\end{figure}

The elliptic fiber used for an  F-theoretic realization of the Trinifcation model is a toric CY-hypersurface in the toric ambient space $\mathbb{P}_{F_{16}}$ given by 
\begin{align}\label{eq:pF16}
p_{F_{16}}=s_1 e_1^2 e_2^2 e_3 e_4 u^3+ s_6 e_1 e_2 e_3 e_4 e_5 e_6 u v w + s_7 e_2 e_3^2 e_6^3 v^2 w + s_9 e_1 e_4^2 e_5^3 v w^2 \, .
\end{align}
The relevant toric data is provided in Figure~\ref{fig:poly16_toric}. The divisor classes on the fiber are 
the hyperplane $H$ and the exceptional divisors $E_1$ to $E_5$. 

In constructing elliptic fibrations $X_{F_{16}}$ with the curve \eqref{eq:pF16} we promote the 
coefficients $s_i$ to sections of the line bundles over the base $B$ given in \eqref{eq:cubicsections}. 
The elliptic fibration of  $X_{F_{16}}$ exhibits three toric sections with two torsional relations among 
them, so that the Mordell-Weil group of the fibration is $\mathbb{Z}_3$ 
\cite{Mayrhofer:2014opa,Klevers:2014bqa}. The fibration admits a  zero section that we choose as
\begin{align}
\hat{s}_0=X_{F_{16}}\cap\{u=0\}&:\quad [0:1:1:s_7:-s_9:1:1:1:1]\,.
\end{align} 
As can be seen by inspecting the Weierstrass model of $X_{F_{16}}$, 
there are three codimension one loci over which the fiber becomes singular, namely $\mathcal{S}_{\SU3^1}=\{s_1=0\}$, $\mathcal{S}_{\SU3^2}=\{s_7=0\}$ and $\mathcal{S}_{\SU3^3}=\{s_9=0\}$. The fiber degenerates to $I_3$ over these three divisors, giving rise to gauge group 
$\SU3^3/\mathbb{Z}_3$ which is characteristic of the Trinification model. The MW-torsion
$\mathbb{Z}_3$ acts simultaneous on the centers of the \SU3 factors. The Cartan divisors for the three $\SU3$ factors read
\begin{align}
\begin{split}
D^{\SU3_1}_{1}=[v]\, , &\quad D^{\SU3_1}_{2}=[w]\,,\\
D^{\SU3_2}_{1}=[e_4]\,,&\quad D^{\SU3_2}_{2}=[s_7]-[e_1]-[e_4]\,, \\
D^{\SU3_3}_{1}=[e_3]\,,&\quad D^{\SU3_3}_{2}=[s_9]-[e_2]-[e_3]\; .
\end{split}
\end{align}

At codimension two we find three loci in $B$, over which the singularity type of the elliptic fibration 
enhances and bifundamental matter is supported,\footnote{The effect of the $\mathbb{Z}_3$ torsion is manifest at codimension two as the only representations which are manifest are singlets under torsion.} see Table~\ref{tab:poly16_matter}.
\begin{table}[H]
\begin{center}
\renewcommand{\arraystretch}{1.2}
\begin{tabular}{|c|c|}\hline
Representation & Locus \\ \hline

$(\three,\overline{\three},\one)$ & $ V(I_{(1)}):= \{ s_1=s_7=0 \}$  \\ \hline

$(\three,\one,\overline{\three})$ & $ V(I_{(2)}):= \{ s_1=s_9=0 \}$  \\ \hline

$(\one,\three, \overline{\three})$ & $ V(I_{(3)}):= \{ s_7=s_9=0 \}$  \\ \hline
\end{tabular}
\caption{\label{tab:poly16_matter}Charged matter representations under $\SU3^3/\mathbb{Z}_3$ and corresponding codimension two loci in $X_{F_{16}}$.}
\end{center}
\end{table}

We complete the base independent analysis of $X_{F_{16}}$ with the computation of its 
second Chern Class as well as its Euler number. Using the methods of \cite{Cvetic:2013qsa}, 
we obtain
\begin{align}
\begin{split}\label{chern16}
c_2(X_{F_{16}})&= -c_1^2 + c_2 - 7 c_1 E_1 + 2 c_1 E_2 - 9 E_4^2 - 2 c_1 E_5 - 2 c_1 E_6 + 4 c_1 H + 3 c_1 \cS_7 + 5 E_1 \cS_7\\
  &\phantom{=} + E_2 \cS_7 + E_3 \cS_7 + 7 E_4 \cS_7 + E_6 \cS_7 - 5 H \cS_7 - 2 c_1 \cS_9 + E_1 \cS_9 - 4 E_2 \cS_9 - 2 E_3 \cS_9\\
   &\phantom{=}- 8 E_4 \cS_9 + E_5 \cS_9 + 4 H \cS_9 - 4 \cS_7 \cS_9 + 4 \cS_9^2 \; ,
\end{split}\\
\begin{split}
\chi(X_{F_{16}})&= 3 (24 c_1^3+4 c_1 c_2-24 c_1^2 \cS_7+8 c_1 \cS_7^2-24 c_1^2 \cS_9+17 c_1 \cS_7 \cS_9-3 \cS_7^2 \cS_9+8 c_1 \cS_9^2\\
		&\phantom{=}-3 \cS_7 \cS_9^2) \label{eq:eulerf16}\; ,
\end{split}
\end{align}
where, as before, $c_1$ and $c_2$ denoted the first and second Chern class of $B$, respectively.

\begin{figure}[ht]
\center
\includegraphics[scale=0.8]{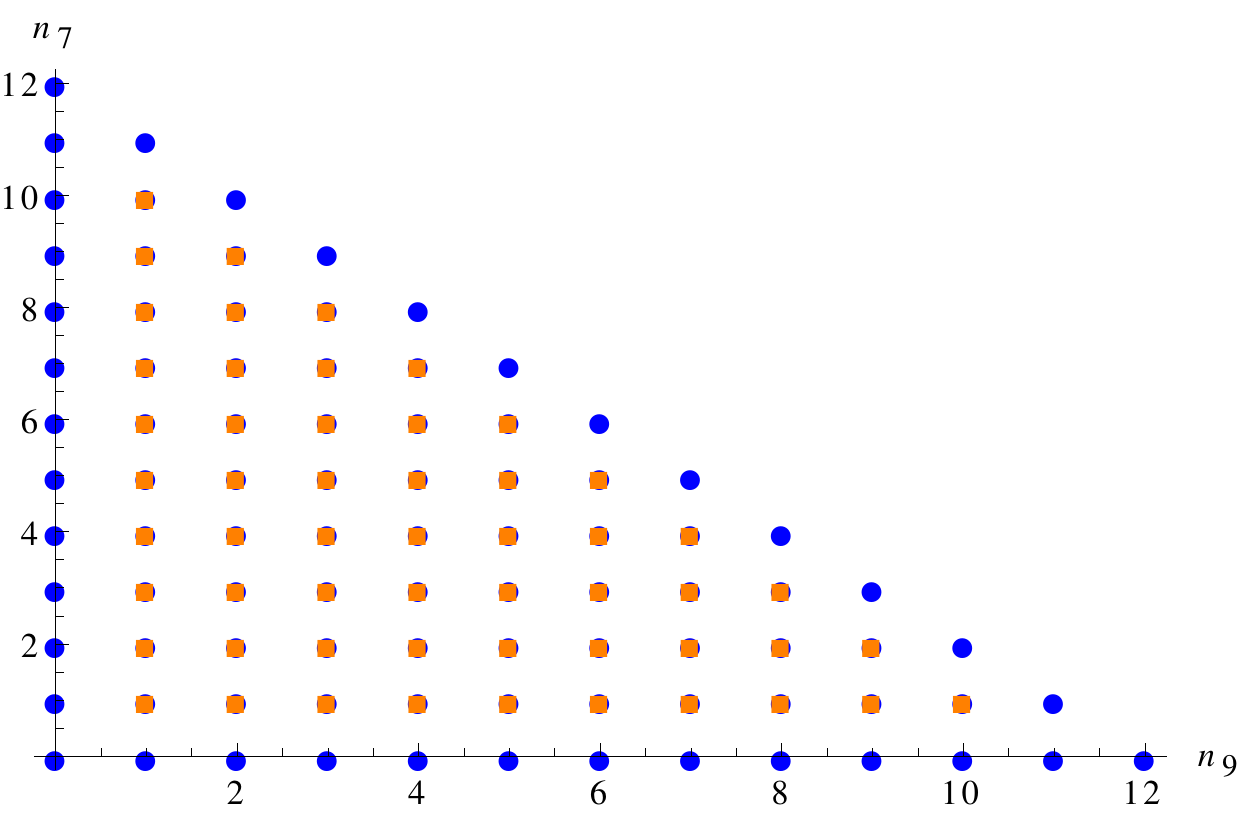}
\caption{\label{fig:f16allowed}The allowed region $(n_7,n_9)$ for the  CY-fourfold $X_{F_{16}}$ for all points that are not on the boundary (orange) allow for $b$ families in the Trinification model.
}
\end{figure}

Next we choose the base $B=\mathbb{P}^3$ and expand the divisors $\cS_7$ and $\cS_9$ in terms 
of the pullback of the hyperplane class $H_B$, c.f.~\eqref{eq:baseP3}. Then we use conditions implied 
by effectiveness of the section $s_i$ in \eqref{eq:cubicsections}, that enter the CY-equation 
\eqref{eq:pF16}, to obtain the allowed region for the pair $(n_7,n_9)$. It is depicted 
in Figure \ref{fig:f16allowed}. 
\subsection{$G_4$-Flux and Chiralities}\label{sec:G4F16}
For the specific base $\mathbb{P}^3$, the full SR-ideal of the ambient space of the CY-fourfold 
$X_{F_{16}}$ reads
\begin{align}
\begin{split} \label{eq:SRF16}
SR&=\{ u e_4, u e_5, uw, uv, ue_6, ue_3, e_1e_5, e_1 w, e_1v, e_1e_6, e_1 e_3, e_1e_2, e_4w, e_4 v, e_4e_6, \\
&\phantom{={}} e_4 e_3, e_4 e_2, e_5v, e_5e_6, e_5 e_3, e_5 e_2, we_6, w e_3, w e_2, ve_3, ve_2, e_6e_2, x_0x_1x_2x_3\} \,.
\end{split}
\end{align}
Here $[u:v:w]$ and the $e_i$, $i=1\ldots,5$ are the homogeneous coordinates on 
$\mathbb{P}_{F_{16}}$ and $x_j$, $j=0,\ldots,3$, denote homogeneous coordinates 
on $\mathbb{P}^3$. As a basis for $H^{(1,1)}(X_{F_{16}})$ we choose
\begin{align}
H^{(1,1)}(X_{F_{16}})&= \langle H_B, S_0, D^{\SU3_1}_1, D^{\SU3_1}_2, D^{\SU3_2}_1, D^{\SU3_2}_2, D^{\SU3_3}_1, D^{\SU3_3}_2 \rangle \, . 
\end{align}

Next, we compute the vertical cohomology ring of $X_{F_{16}}$ as described in Section 
\ref{sec:review} as a quotient ring using the SR-ideal \eqref{eq:SRF16} together 
with the following intersection numbers, that descend from the toric intersections in 
$\mathbb{P}_{F_{16}}$:
\begin{align}
\begin{split}
H_B^3\cdot S_0^2 = -1 &,\quad H_B^3\cdot (D^{\SU3_1}_2)^2 = -2\, ,\\
H_B^3\cdot (D^{\SU3_2}_2)^2 = -2 &,\quad  H_B^3\cdot (D^{\SU3_3}_2)^2 = -2\, .
\end{split}
\end{align}
With this information at hand we can obtain all quartic intersections in $X_{F_{16}}$. The dimension of 
$H^{(2,2)}_V(X_{F_{16}})$ is found after taking all possible products of two elements in $H^{(1,1)}$ and evaluating the rank of their inner product matrix, yielding 
\begin{align}
{\rm dim}(H^{(2,2)}_V(X_{F_{16}}))&=9\; .
\end{align}
As a basis for $H^{(2,2)}_V(X_{F_{16}})$ we  choose
\begin{align}
\begin{split}
H^{(2,2)}_V(X_{F_{16}})&= \langle H_B^2, H_B S_0, D^{\SU3_1}_1 H_B, D^{\SU3_1}_2 H_B, D^{\SU3_2}_1 H_B, D^{\SU3_2}_2 H_B, \\
		   &\phantom{= \langle a} D^{\SU3_3}_1 H_B, D^{\SU3_3}_2 H_B,S_0^2 \rangle \; ,
\end{split}
\end{align}
which we use to make an ansatz for the most general  $G_4$-flux. Since the $G_4$-flux must be 
consistent with the matching of M- and F-theoretical CS terms, we have to impose the constraints 
\eqref{eq:csfluxconstraints}, which amount to eight independent constrains. We are left with the 
following one parameter $G_4$-flux:
\begin{align}\label{eq:fluxf16}
\begin{split}
G_4 &= a_9\big [\!\! -\tfrac13 H_B\cdot( n_7 D^{\SU3_3}_1 \!\!  +2n_7D^{\SU3_3}_2  \!\! + n_9 D^{\SU3_2}_1  \!\!+ 2n_9 D^{\SU3_2}_2)  \\
	&\phantom{= a_9\big [\!\! }+ n_7 n_9H_B^2  + 4 H_B\cdot S_0 + S_0^2 \big ] \; .
\end{split}
\end{align}
The parameter $a_9$ must be consistently quantized so that  the $G_4$-flux quantization condition 
\eqref{eq:quantization} with $c_2(X_{F_{16}})$ as given in \eqref{chern16} is obeyed.  Again,
we ensure the quantization indirectly by choosing $a_9$ so that the number $n_{\rm D3}$
is a positive integer and that all  3D CS-terms \eqref{eq:thetaM} are integral. This issue is discussed 
below.

As a first step towards the computation of the 4D matter chiralities we provide the homology classes 
for the  matter surfaces. Using the results from Table \ref{tab:poly16_matter}, they read
\begin{align}\label{eq:mattersurfF16}
\begin{split}
\mathcal{C}^w_{(\three,\overline{\three},\one)}&=	\cS_7 (3 [K_B^{-1}] - \cS_7 - \cS_9) (H-E_1 + \cS_9)\;,\\
\mathcal{C}^w_{(\three,\one,\overline{\three})}&=	\cS_9 (3 [K_B^{-1}] - \cS_7 - \cS_9) (H-E_2 + \cS_9)\;,\\
\mathcal{C}^w_{(\one,\three,\overline{\three})}&=	\cS_7 \cS_9 (2H-E_1 - E_2 - E_3 - E_4 + K_B^{-1}] - \cS_7 + \cS_9)\;,
\end{split}
\end{align}
where for each matter surface we have chosen an node in the fiber at codimension two with weight 
$w$ of the respective representation which is not intersected by the zero section. Integrating 
the $G_4$-flux over the  surfaces \eqref{eq:mattersurfF16} leads according to \eqref{eq:g4chiralities} 
the following chiral indices:
\begin{align}\label{chiF16}
\chi_{(\three,\overline{\three},\one)}
= \chi_{(\three,\one,\overline{\three})}
= \chi_{(\one,\three,\overline{\three})}&=	\tfrac13 n_7 n_9 (-12 + n_7 + n_9) a_9						\; .
\end{align}
Note that all chiralities are equal, as expected in order for the cubic \SU3 anomalies to cancel. 
We also emphasize that at the boundary of the allowed region (see Figure \ref{fig:f16allowed}) we can 
not have a chiral theory, as all chiralities \eqref{chiF16} vanish there. 

As before, we express the parameter $a_9$ in terms of the number of families $b\equiv \chi_{(\three,\overline{\three},\one)}$ as
\begin{align}\label{eq:familyparameterf16}
a_9 = \frac{3b}{n_7 n_9 \left(-12 + n_7 + n_9  \right)} \; .
\end{align}
For all allowed values of $(n_7,n_9)$,  we explore which positive integral values for $b$ lead
to a canceled  D3-brane tadpole with a positive  integral number $n_{\rm D3}$ of D3-branes without
adding horizontal $G_4$-flux. As shown in Table \ref{fig:f16allowed}, we find that for three families 
($b=3$), which is also the minimal value of families, the D3-brane tadpole is canceled at nine different 
values for $(n_7,n_9)$.  We have also checked for the three-family models, that 
all 3D CS-terms \eqref{eq:thetaM} are integral. We observe that Table \ref{fig:f16allowed} is symmetric 
as expected by the symmetries of polyhedron  $F_{16}$ and that there is a family structure for every
allowed value of $(n_7,n_9)$.
\begin{table}[t!]
\begin{center}
\footnotesize
\begin{tabular}{c|cccccccccc}
{\large $_{n_7} \backslash ^{n_9}$} & 1 & 2 & 3 & 4 & 5 & 6 & 7 & 8 & 9 & 10\\ \hline
10 & $( 5;120 )$ &  &  &  &  &  &  &  &  &  \\
9 & {\color{red}$( 3;94 )$} & {\color{red}$( 3;94 )$} &  &  &  &  &  &  &  & \\
8 & $( 4;72 )$ & $( 8;69 )$ & $( 4;72 )$ &  &  &  &  &  &  &  \\
7 & $( 14;48 )$ & $( 7;54 )$ & $( 7;54 )$ & $( 14;48 )$ &  &  &  &  &  &  \\
6 & $( 5;50 )$ & $( 8;44 )$ & {\color{red}$( 3;44 )$} & $( 8;44 )$ & $( 5;50 )$ &  &  &  &  &  \\
5 & $( 5;50 )$ & $( 5;42 )$ & $( 10;36 )$ & $( 10;36 )$ & $( 5;42 )$ & $( 5;50 )$ &  &  &  &  \\
4 & $( 14;48 )$ & $( 8;44 )$ & $( 10;36 )$ & $( 16;30 )$ & $( 10;36 )$ & $( 8;44 )$ & $( 14;48 )$ &  &  &  \\\
3 & $( 4;72 )$ & $( 7;54 )$ & {\color{red}$( 3;44 )$} & $( 10;36 )$ & $( 10;36 )$ & {\color{red}$( 3;44 )$} & $( 7;54 )$ & $( 4;72 )$ &  &  \\
2 & {\color{red}$( 3;94 )$} & $( 8;69 )$ & $( 7;54 )$ & $( 8;44 )$ & $( 5;42 )$ & $( 8;44 )$ & $( 7;54 )$ & $( 8;69 )$ & {\color{red}$( 3;94 )$} & \\
1 & $( 5;120 )$ & {\color{red}$( 3;94 )$} & $( 4;72 )$ & $( 14;48 )$ & $( 5;50 )$ & $( 5;50 )$ & $( 14;48 )$ & $( 4;72 )$ & {\color{red}$( 3; 94 )$} & $( 5; 120 )$ \\
\end{tabular}
\caption{\label{tab:families_f16}The entries $(b;n_{\rm D3})$ show the minimal number of families $b$ for which the number of D3 branes $n_{\rm D3}$ is integral and the 3D CS-terms are quantized.}
\end{center}
\end{table}

As a cross-check of our results \eqref{chiF16} we verify the matching \eqref{eq:M=F} of CS-terms in 
F- and M-theory. The non-vanishing CS-terms on the F-theory side computed using \eqref{eq:thetaF}
read
\begin{align}
\begin{split}
\Theta^\mathrm{F}_{11} &= -(\chi_{(\three,\overline{\three},\one)}+\chi_{(\three,\one,\overline{\three})})\;, \qquad
\Theta^\mathrm{F}_{12} = \tfrac12 (\chi_{(\three,\overline{\three},\one)}+\chi_{(\three,\one,\overline{\three})})\;, \qquad
\Theta^\mathrm{F}_{22} = -(\chi_{(\three,\overline{\three},\one)}+\chi_{(\three,\one,\overline{\three})})\;,\\
\Theta^\mathrm{F}_{33} &= -2 \chi_{(\one,\three,\overline{\three})} \;, \qquad\qquad\qquad\,\,
\Theta^\mathrm{F}_{34} = \tfrac12 (3 \chi_{(\three,\overline{\three},\one)} - \chi_{(\one, \three, \overline{\three})}) \;, \quad\,\,\,
\Theta^\mathrm{F}_{44} = -(3 \chi_{(\three,\overline{\three},\one)} - \chi_{(\one, \three, \overline{\three})})\;,\\
\Theta^\mathrm{F}_{55} &= -(3 \chi_{(\three,\overline{\three},\one)} - \chi_{(\one, \three, \overline{\three})}) \;, \quad\,\,\,
\Theta^\mathrm{F}_{56} = \tfrac12 (3 \chi_{(\three,\overline{\three},\one)} - \chi_{(\one, \three, \overline{\three})}) \;, \quad\,\,\,
\Theta^\mathrm{F}_{66} = -2\chi_{(\one,\three,\overline{\three})} \; .
\end{split}
\end{align}
We readily compute the CS-terms \eqref{eq:thetaM} in M-theory, which allows us to reproduce 
precisely the chiralities in \eqref{chiF16}.

\subsection{Phenomenological Discussion}\label{sec:phenoF16}

The breaking from the Trinification model to the SM proceeds via two successive Higgsings. Geometrically, the Higgsings
correspond to blow-downs in $X_{F_{16}}$ induced by toric blow-downs in the ambient space $\mathbb{P}_{F_{16}}$
of the elliptic fiber. Thus, we can geometrically visualize the Higgsing directly in the toric diagram of $F_{16}$, see 
\cite{Klevers:2014bqa} for details.

More concretely, in order to obtain the CY-hypersurface $X_{F_{11}}$ starting from $X_{F_{16}}$, we have to perform two 
blow-downs in the fiber of $X_{F_{16}}$ that are identified by requiring that the fiber polyhedron $F_{16}$ is mapped to $F_{11}$. 
There are three possible ways to achieve this.  However all of these are all equivalent due to the symmetries of the 
polyhedron. For concreteness we choose here the transition $X_{F_{16}}\rightarrow X_{F_{11}}$ induced 
by blowing down the divisor $e_5=0$ and subsequently the divisor $e_4=0$ in the toric diagram of $F_{16}$, see Figure 
\ref{fig:poly16_toric}. Note that after the blow downs we indeed get the toric diagram of $F_{11}$ in Figure \ref{fig:poly11_toric} 
reflected along the horizontal axis passing through the origin.

The blow-down process restricts the allowed region in Figure \ref{fig:f16allowed} of $X_{F_{16}}$ due to the requirement 
that the divisors associated to the sections $s_3$ and $s_5$, which are part of \eqref{eq:pF11} but not of \eqref{eq:pF16}, 
are effective. This removes all models which lie above the line $4+n_7-n_9=0$ in Figure \ref{fig:f16allowed}. Once we exclude 
these un-Higgsable models, we can compare the number of generations in $X_{F_{16}}$ and $X_{F_{11}}$ in 
Figures \ref{tab:families_f11} and \ref{tab:families_f16}. In order to perform such a comparison we have to bear in mind that
the polyhedron obtained by Higgsing $X_{F_{16}}$  and the polyhedron specifying  $X_{F_{11}}$ are related by a reflection, as 
mentioned above. Thus, one has to perform a redefinition of the integers $(n_7,n_9)$ before and after Higgsing 
which amounts to the shift $(n_7,n_9)\rightarrow (8-n_7,n_9)$. Under this map we see that the point $(n_7,n_9)=(3,6)$ in 
$X_{F_{16}}$ maps to $(5,6)$ in $X_{F_{11}}$, both of supporting models with $b=3$. This suggests that there is a 
toric Higgsing of a Trinification model to a  SM with three families.  We take this as the example for the following
phenomenological discussion.

On the field theory side the described transition $X_{F_{16}}\rightarrow X_{F_{11}}$ proceeds by a VEV of the field in 
the  representation $(\three,\overline{\three},\one)$ at the matter curve $s_1=s_7=0$, cf.~Table \ref{tab:poly16_matter}. For 
reasons that will become clear in the following, the Higgsing down to the MSSM is only possible if one has in addition to the 
three chiral families  also has two vector-like pairs $(\three,\overline{\three},\one)$, $(\overline{\three},\three,\one)$ 
\cite{Lazarides:1993sn}. While one pair, which we denote as $H^1$, $\bar{H}^1$ is needed for the intermediate breaking 
$\SU3^3/\mathbb{Z}_3\rightarrow \SU{3}\times\SU{2}^2\times \U1$, the second pair $H^2$, $\bar{H}^2$ is needed because 
the representation, which breaks the $\SU2\times \U1$ down to the hypercharge generator, is contained in the 
$(\three,\overline{\three},\one)$ representation, too. Therefore, in addition to the three generations of chiral fields 
$(\lambda_i,\mathcal{Q}_i,\bar{\mathcal{Q}}_i)$, we assume that our model allows for two massless vector-like pairs of Higgs fields $H^1$, $\bar{H}^1$ and $H^2$, $\bar{H}^2$. 

In the first Higgsing inducing the symmetry breaking $\SU3^3/\mathbb{Z}_3\rightarrow \SU{3}\times\SU{2}^2\times \U1$ we 
arrive at the following decomposition of fields composing the chiral families of our model:
\begin{align}\label{eq:familienF13}
\begin{split}
\lambda_i=(\three,\overline{\three},\one)&\rightarrow (\two,\two,\one)_{0}+(\one,\two,\one)_{-3}+(\two,\one,\one)_3+(\one,\one,\one)_0\;,\\
\mathcal{Q}_i=(\overline{\three},\one,\three)&\rightarrow(\two,\one,\three)_{-1}+(\one,\one,\three)_{+2}\;,\\
\bar{\mathcal{Q}}_i=(\one,\three,\overline{\three})&\rightarrow(\one,\two,\overline{\three})_{1}+(\one,\one,\overline{\three})_{-2}\;.\\
\end{split}
\end{align}
Similarly the Higgses of the Trinification model decompose as
\begin{align}
\begin{split}\label{eq:HiggsesF13}
H^1,H^2=(\three,\overline{\three},\one)&\rightarrow (\two,\two,\one)_{0}+(\one,\two,\one)_{-3}+(\two,\one,\one)_{+3}+(\one,\one,\one)_0\;,\\
\bar{H}^1,\bar{H}^2=(\overline{\three},\three,\one)&\rightarrow (\two,\two,\one)_{0}+(\one,\two,\one)_{+3}+(\two,\one,\one)_{-3}+(\one,\one,\one)_0\;.
\end{split}
\end{align}
Here, the generator of the unbroken $\U1$ is given by
\begin{equation}
Q=T^8_1+T^8_2\,,\quad\text{with}\quad T^8_{1,2}=\text{diag}(1,1,-2)\,,
\end{equation}
where the subscripts $1$, $2$ label the corresponding $\SU3$ factors. 

Next we need to break one of the \SU2 factors together with the $\U1$-factor to the hypercharge $\U1_Y$. 
To this end, one of the fields in either the representation 
$(\one,\two,\one)_{-3}$ or $(\two,\one,\one)_{+3}$ in the decomposition of the representation
$(\three,\overline{\three},\one)$ has to pick a VEV. Note also that for the pair $H^1$, $\bar{H}^1$, both of these fields (and their 
complex conjugates) will be absorbed as the longitudinal modes of the massive vector fields. Hence, we have to pick the fields 
for the second Higgsing as irreducible representations stemming from the decomposition of $H^2$ and $\bar{H}^2$ in 
\eqref{eq:HiggsesF13}. For concreteness we take the states $(\one,\two,\one)_{-3}\subset H^2$, $(\one,\two,\one)_{+3}\subset \bar{H}^2$ to be responsible for the second breaking. In that case the hypercharge generator written in terms of the $\U1$ generator $Q$ and the Cartan generator $T^3$ of the first $\SU2$  reads
 \begin{equation}
Q_Y=-(T^3+Q/6)\,.
\end{equation}
The decomposition of the chiral matter representations in terms of the SM gauge group is given in Table 
\ref{tab:Trinidecomposition}. Here we immediately see that each family at the Trinification level provides an entire SM family, 
extended by two right handed neutrinos $\bar{\nu}_{1,i}$, $\bar{\nu}_{2,i}$, a vector-like pair of color triplets $D_i$, $\bar{D}_i$ 
and a pair of SM-like Higgs fields $H_{u,i}$, $H_{d,j}$.
\begin{table}[H]
\begin{center}
\renewcommand{\arraystretch}{1.2}
\begin{tabular}{|ccl|} \hline
Tri-Rep &  & SM decomposition \\ \hline
$\lambda_i$ & $\rightarrow$ & $H_{d,i}$ : $(\one,\two)_{-1/2}$ , $H_{u,i}$ : $(\one,\two)_{1/2}$ , $\overline{e}_i$ : $(\one,\one)_{1}$ , \\
&& $L_i$ :  $(\one,\two)_{-1/2}$, $\bar{\nu}_i^1$ : $(\one,\one)_0$ , $\bar{\nu}_i^2$ : $(\one,\one)_0$\\
$\mathcal{Q}_i$ & $\rightarrow$ & $Q_i$: $(\three,\two)_{\frac16}$ ,  $D_{i}$ : $(\overline{\three},\one)_{-\frac13}$ \\
$\bar{\mathcal{Q}}_i$   & $\rightarrow$ &$\overline{u}$ : $(\overline{\three},\one)_{-\frac23}$ , $\overline{d}_i$ : $(\overline{\three},\one)_{\frac13}$ , $\overline{D}_{i}$ :  $(\overline{\three},\one)_{\frac13}$ \\ \hline
$H^1$, $\bar{H}^1$ & $\rightarrow$ & $H_{d}^1$, $\bar{H}_{u}^1$ : $(\one,\two)_{-1/2}$ , $H_{u}^1$, $\bar{H}_{d}^1$ : $(\one,\two)_{1/2}$ \\
$H^2$, $\bar{H}^2$ & $\rightarrow$ & $H_{d}^2$, $\bar{H}_{u}^2$, $H_d^{2\prime}$ : $(\one,\two)_{-1/2}$ , $H_{u}^1$, $\bar{H}_{d}^1$, $H_d^{2\prime}$ : $(\one,\two)_{1/2}$ \\\hline
\end{tabular}
 \caption{\label{tab:Trinidecomposition}The Trinification representations decomposed in terms of their MSSM constituents. We have also included the charged fields inside of the fields $H^1$, $H^2$ which do not participate in the Higgs process.}
\end{center}
\end{table}
\begin{table}[htb!]
\begin{center}
\renewcommand{\arraystretch}{1.3}
\begin{tabular}{|c|c|}\hline
Yukawa & Locus \\ \hline
$(\three,\overline{\three},\one) \cdot \overline{(\three,\one,\overline{\three})} \cdot (\one,\three,\overline{\three})$ & $s_1=s_7=s_9=0$ \\ \hline
$(\three,\overline{\three},\one) \cdot (\three,\overline{\three},\one) \cdot (\three,\overline{\three},\one)$ & $s_1=s_6=s_7=0$\\ \hline
$(\three,\one,\overline{\three}) \cdot (\three,\one,\overline{\three}) \cdot (\three,\one,\overline{\three})$ & $s_1=s_6=s_8=0$\\ \hline
$(\one,\three,\overline{\three}) \cdot (\one,\three,\overline{\three}) \cdot (\one,\three,\overline{\three})$ & $s_6=s_7=s_9=0$\\ \hline
\end{tabular}
\caption{\label{tab:poly16_yukawa}Codimension three loci and corresponding Yukawa points for $X_{F_{16}}$. }
\end{center}
\end{table}
For the decoupling of the exotics and the discussion of the couplings of the fields, we have to refer to the geometrically allowed 
Yukawa couplings in $X_{F_{16}}$ which are given in Table \ref{tab:poly16_yukawa} \cite{Klevers:2014bqa}. 
First notice that in general one has two scales $\Lambda_1$ and $\Lambda_2$ associated to each of the two symmetry 
breakings from Trinification to SM. However, radiative corrections will push these scales towards each other, and due to that we 
can simply assume that both Higgsings occur simultaneously at some scale $\Lambda$. Given the allowed couplings 
$H^1\mathcal{Q}_i\bar{\mathcal{Q}}_i$ and $H^2\mathcal{Q}_i\bar{\mathcal{Q}}_j$ we see that the exotic triplets $D_i$, 
$\bar{D}_i$ pick up a mass of order $\Lambda$. We also note from Table \ref{tab:Trinidecomposition} that we get eight pairs of 
SM-like Higgses. Nevertheless, working out all bilinears among them, which result from the three point couplings of the 
Trinification model with VEV insertions, one sees that essentially all of the Higgses get lifted at the Trinification scale 
$\Lambda$. One might hope that this can be fixed by tuning the complex structure of $X_{F_{16}}$ to suppress the 
corresponding Yukawa couplings, however, we have to recall that there are only two structurally different types of Yukawa 
couplings. Indeed,  on the one hand we have a three point coupling involving a single matter curve. 
On the other hand we have a coupling involving the three different matter curves, which if suppressed, will imply that Yukawa couplings for quarks and leptons are suppressed.

We conclude with another remark regarding the presence of vector-like pairs in the model.
As we discussed above, we need vector-like matter $H_1$, $\bar{H}_1$, $H_2$ and $\bar{H}_2$  
in the representations $(\overline{\three},\three,\one)$ and $(\three,\overline{\three},\one)$, respectively, for the Higgsing down to the 
MSSM. It is then natural to expect also additional vector-like matter in the representations $(\overline{\three},\one,\three)$ and 
$(\one,\three,\overline{\three})$, which would give rise to exotics beyond those presented in Table 
\ref{tab:Trinidecomposition}.\footnote{Again, it is expected that these exotics acquire masses of order $\Lambda$, too.} 
Indeed, this can be motivated geometrically by the $SL(2,\mathbb{Z})$-symmetry of the polyhedron $F_{16}$ that acts as a 
permutation symmetry on the three SU(3) gauge factors and corresponding representations. We have seen that this symmetry 
is realized on the chiral spectrum in Table \ref{tab:families_f16} and it is natural to expect that it also holds for the vector-like 
sector of the theory. In addition, this permutation symmetry will also be responsible for the unification of the SM gauge 
couplings above the Trinification scale \cite{Dvali:1994wj}. Furthermore, in  \cite{Lazarides:1993sn,Dvali:1996fc} 
the possibility of having a light SM Higgs pair has been related to the presence of additional 
discrete $R$ and non-$R$ symmetries in Trinification models with vector-like pairs. These 
symmetries could reduce to the standard matter parity after the Trinification group is broken down to the SM gauge group, 
and hence they could prevent the model from an exceedingly fast proton decay. It would be interesting
to find ways to realize these additional symmetries in F-theory.


\section{Conclusions}
\label{sec:conclusion}

In this work we have presented explicit, globally consistent four-dimensional F-theory compactifications that have the standard 
model gauge group and three chiral families. Moreover we considered embeddings of the standard model into the Pati-Salam 
or trinification models using a toric realization of the relevant Higgs effects.

The models considered in this work result from F-theory compactifications on Calabi-Yau fourfolds, with their elliptic fibers 
defined as toric hypersufaces in the three 2D toric ambient spaces  $\mathbb{P}_{F_{11}}$, $\mathbb{P}_{F_{13}}$ and 
$\mathbb{P}_{F_{16}}$, see e.g.~\cite{Klevers:2014bqa}. While the gauge symmetry, the type of matter representations and 
the possible Yukawa couplings are properties independent of the choice for the base manifold, we fix the base manifold to be 
$B=\mathbb{P}^3$ in order to construct the vertical cohomology in each case. This allows us to find explicit solutions for the 
chirality-inducing $G_4$-flux and ultimately, to determine the possible 4D chiralities of matter fields. The expressions for the 
$G_4$-flux in each type of compactification are shown to satisfy the consistency conditions imposed by the matching of F- and 
M-theoretical Chern-Simons terms, as well a the 
D3-brane tadpole cancelation condition with a positive, integral number of D3-branes. 
The $G_4$-flux quantization condition is ensured indirectly  by the analogous quantization condition of the 3D CS-terms. We 
also show that the obtained chiralities are consistent with field theoretic calculations of CS-terms and anomalies.

In this work we have only considered the vertical part of the middle cohomology, which is responsible for the generation of 
chirality. With no $G_4$-flux along the horizontal components we observe that the D3-brane tadpole cancellation imposes very 
stringent constraints on the minimal number of generations permitted in each of the models, which happens to be precisely 
three. We explicitly identify the  consistent  three-family models in our list of concrete models. 
It is expected that the addition of horizontal $G_4$-flux ameliorates these bounds and permits a larger amount of consistent 
three-family solutions in the discrete set of considered CY-fourfolds. 

For the Pati-Salam and Trinification models we also described the Higgs transition to the Standard Model. On the geometric 
side, effectiveness conditions of the divisor classes needed to specify the models after the transition determine points in the 
allowed region for $(n_7,n_9)$ of $X_{F_{13}}$ and $X_{F_{16}}$, for which the Higgsing is possible. While in the 
six-dimensional case, these conditions are known to ensure a  D-flat Higgsing, in four dimensions, a similar field theoretical 
understanding remains elusive and requires a better understanding of vector-like matter in F-theory, beyond chiral indices. 
Indeed, we presume that for values of $(n_7,n_9)$, where some divisors fail to be effective, one does not have a light pair of 
vector-like fields to carry out the Higgsing. Due to the current lack of control of the vector-like sector of F-theory, 
we simply assume that over the points where all divisors defining the model remain effective after the transition 
one has the necessary vector-like pairs of Higgses and that the transition is indeed possible. Since the supersymmetric Higgs 
mechanism does not change the net chiralities, we expect that a model based on $X_{F_{13}}$ or $X_{F_{16}}$ with 
three families at a given point $(n_7,n_9)$ maps to a model with three families in $X_{F_{11}}$, at the same point 
$(n_7,n_9)$. Remarkably for both $X_{F_{13}}$ and $X_{F_{16}}$ we find a point with three families before and 
after Higgsing. If we in addition require that the number of D3-branes remains constant in the transition, 
we must add horizontal  $G_4$-flux after the Higgsing in order to compensate for the change in the Euler number of the 
CY-fourfold. The systematic inclusion of horizontal $G_4$-flux and their effect on the Higgsing, as well as the connection to
$G_4$-flux quantization, would be interesting to study in future works.

In this work we have also made some remarks on the phenomenology of the studied models. A rough look at the 
superpotential at quadratic and cubic order of the resulting effective field theories lead to the well known observation that the 
bare models have some phenomenologically unapealing features, such as the prediction of a fast proton decay or the
difficulty to retain a light pair of electroweak Higgs fields in the spectrum. However, it is likely that the proton decay operators 
can be kept under control if one goes to special points in complex structure moduli space of the CY-fourfolds. 
On the field theory side this usually corresponds to the existence of an accidental discrete symmetry.
It would be very interesting whether there exists horizontal $G_4$-flux that stabilizes the complex structure of the
CY-fourfold at these points in moduli, following the general arguments discussed in \cite{Bizet:2014uua}. 

Finally, we have also computed the Hodge numbers of the considered CY-fourfolds with three-families. For our simple
choice of base $\mathbb{P}^3$ we always have $h^{(2,1)}=0$ which constraints cosmological F-theory applications of our 
models. Clearly, it is an interesting future direction to extend the phenomenological analysis carried out in this work to other 
bases $B$ allowing for richer possible applications to cosmology.

\subsection*{Acknowledgments}

We would like to thank Hans Jockers, Albrecht Klemm
and Hernan Piragua  for helpful discussions and comments.
M.C. and D.K. would like to thank Paul Langacker for discussions and collaboration on related topics.
D.K.~thanks the Bethe Center for Theoretical Physics Bonn for hospitality during completion of the
project. D.M.~thanks the DESY theory group, where part of this work was completed, for hospitality. 
The work of M.C.~is supported by the DOE grant DE-SC0007901, the Fay R. and Eugene L. Langberg 
Endowed Chairand the Slovenian Research Agency (ARRS).  
The work of D.M. was partially supported by the Deutsche Forschungsgemeinschaft under the Collaborative Research
Center (SFB) 676 Particles, Strings and the Early Universe.
The work of P.O.~and J.R.~is partially supported by a scholarship of the Bonn-Cologne Graduate School
BCGS, the SFB-Transregio TR33 The Dark Universe (Deutsche Forschungsgemeinschaft) and the
European Union 7th network program Unification in the LHC era (PITN-GA-2009-237920).

\appendix

\section{Hodge Numbers of CY-Fourfolds}
\label{app:HodgeNumbers}

In this appendix we discuss the computation of Hodge number of CY-fourfolds given as toric 
hypersurfaces. We readily use these methods for the calculation of the Hodge numbers of the three 
families of CY-fourfolds $X_{F_{11}}$, $X_{F_{13}}$ and $X_{F_{16}}$. 
 
A CY-fourfold has the independent Hodge numbers $h^{(1,1)}, h^{(2,1)}$ and $h^{(3,1)}$.
They are related to the Euler number as
\begin{align}
\label{eq:EulerHodge}
\chi(X) = 6 \left( 8 + h^{(1,1)} + h^{(3,1)} - h^{(2,1)}      \right) \, .
\end{align}
Since one can compute $\chi(X)$ independently by integrating the top Chern class $c_4(X)$ over $X$,
it is sufficient to calculate $h^{(2,1)}$ and $h^{(3,1)}$ to obtain all independent Hodge numbers.

For a CY-fourfold given as a toric hypersurface specified by a pair of dual five dimensional reflexive 
lattice polyhedra $(\Delta,\Delta^*)$, one can  use the combinatorial Batyrev formulas 
(see \cite{Klemm:1996ts} and references therein) to calculate the Hodge numbers as
\begin{align} \label{eq:HodgeToric}
\begin{split}
h^{(1,1)} &= l(\Delta) - (4+2) - \sum_{\text{dim} \Theta = 4} l^\prime (\Theta) + \sum_{\text{codim} \Theta_i = 2 } 	l^\prime (\Theta_i) l^\prime (\Theta^*_i) \, ,\\
 h^{(3,1)} &= l(\Delta^*) - (4+2) -\sum_{\dim \Theta^* =4} l'(\Theta^*)+
              \sum_{\text{codim}  \Theta_i =2} l'(\Theta_i) l'( \Theta_i^*)\, ,\\
h^{(2,1)} &= \sum_{\text{codim} \Theta_i = 3} l^\prime (\Theta_i) l^\prime (\Theta^*_i) \, .
\end{split}
\end{align}
Here, $\Theta$ ($\Theta^*$) denote faces of $\Delta$ ($\Delta^*$), while the sum is over pairs 
$(\Theta_i,\Theta^*_i)$
of dual faces. 
The $l(\Theta)$ and $l'(\Theta)$ count the total number of integral points of a 
face $\Theta$ and the number inside the face $\Theta$, respectively.
Finally, $l(\Delta)$ is 
the total number of integral points of the polyhedron $\Delta$.

In the following we are considering fourfolds that have the elliptic curves in $\mathbb{P}_{F_{11}}$ ,$\mathbb{P}_{F_{13}}$ and 
$\mathbb{P}_{F_{16}}$ as a fiber over a $\mathbb{P}^3$ base space. An explicit expression for the total polyhedra
of the 5D toric ambient space of these fibrations can be found in Appendix \ref{app:polytope}. We only need some general 
observations about these polyhedra here. The polytope of the $\mathbb{P}^3$ base always contributes four points. Hence we 
can write
\begin{align}
l(\Delta) = 4 + \#\text{Points}(F_i)  \, .
\end{align}
For all fibrations of this type we find that there are never points within codimension two and three facets of the polyhedra i.e.
\begin{align}
l^\prime (\Theta_{\text{codim=2},i}) = 0 \quad (\forall \, i) \, , \qquad
l^\prime (\Theta_{\text{codim=3},i}) = 0 \quad( \forall \, i) \, .
\end{align}
Furthermore we use the following observation that we made in \cite{Klevers:2014bqa}, namely
\begin{align}
\# \text{Points}(F_i)-4 = \text{Rank}(G_{F_{i}}) \, ,
\end{align}
at a {\it generic} stratum of the fibration, i.e. on a bulk stratum of the allowed region.
Hence the above formula \eqref{eq:HodgeToric} for the Hodge number $h^{(1,1)}$ simplifies to
\begin{align}
h^{(1,1)} = \#\text{Points}(F_i)- 2 - \sum_{\text{dim} \Theta = 4} l^\prime (\Theta) = \text{Rank}(G_{F_{i}})+2 \, .
\end{align}
Note that, in contrast to this formular,  the formula \eqref{eq:HodgeToric} 
is valid for all strata of the fibration even at the boundary of the allowed regions were divisors are 
switched off and the rank of the gauge group is reduced.
This rank reduction is precisely taken care of by vertices that move into the interior of dimension four facets and therefore
correct the above formula for the $h^{(1,1)}$ numbers.

Similarly we find, that for all of our models
\begin{align}
h^{(2,1)} = 0 \, .
\end{align}
Using \eqref{eq:EulerHodge} we can thus give a closed formula for all $h^{(3,1)}$
\begin{align}
h^{(3,1)} = \frac{\chi}{6} - 10 - \text{Rank}(G_{F_i}) \, .
\end{align}
Furthermore, by having fixed the above Hodge numbers we can obtain the Hodge number  $h^{(2,2)}$  as well using
\begin{align}
\begin{split}
h^{(2,2)}= 2 \left(22 + 2h^{(1,1)} + 2 h^{(3,1)} - h^{(2,1)}       \right) 
= 2 \left( 6+\frac{\chi}{3} \right) \, .
\end{split}
\end{align}
By using the explicit presentation of the vertical cohomology ring $H^{(2,2)}_V(X_{F_i})$ as a quotient ring, we can compute 
$h^{(2,2)}_V$.  With the knowledge of the full $h^{(2,2)}$ we can compute the dimension of the horizontal cohomology.

We note that the absence of any $(2,1)$-forms is of particular interest for cosmological applications as the resulting 
three-forms from could  be used to obtain axions to drive inflation as discussed in an F-theory context in 
\cite{Grimm:2014vva,Garcia-Etxebarria:2014wla}. However these specific type of axions are absent in all of our models. We conclude that the $\mathbb{P}^3$ base is too simple to allow for these features.\footnote{We still find generically a lot of complex structure moduli $h^{(3,1)}$ that can be used as well. However the specific scenario in \cite{Grimm:2014vva} is excluded.}

\section{Concrete Toric Lattice Polyhedra of 4D Chiral Models}
\label{app:polytope}
Using the algorithm of \cite{Braun:2013nqa} we can construct the CY-manifolds $X_{F_i}$ as concrete toric hypersurfaces 
associated to a five dimensional lattice polyhedron that
specifies the underlying 5D ambient space. In our case the ambient space is a $\mathbb{P}_{F_i}$-fibration with a $\mathbb{P}^3$ base specified by two numbers $n_7$ and $n_9$.
The polyhedron is given in Table \ref{tab:5Dpolytope}.
\begin{table}[H]
\begin{center}
\begin{tabular}{|c|ccc|cc|}\hline
variable & \multicolumn{3}{c|}{Base Vertex} & \multicolumn{2}{c|}{Fiber Vertex} \\ \hline
$z_1$ & 1 & 1 & 1 & $n_9-4$ & $4-n_7$ \\
$z_2$& -1 & 0 & 0 & 0 & 0 \\
$z_4$& 0 & -1 & 0 & 0 & 0 \\
$z_3$& 0 & 0 & -1 & 0 & 0 \\ \hline
$\hat{v}_0$& 0 & 0 & 0 & 0 & 0 \\
$\hat{v}_i$& 0 & 0 & 0 &  \multicolumn{2}{c|}{$v_i$} \\ \hline
\end{tabular}
\caption{\label{tab:5Dpolytope}The five dimensional polyhedron describing a $\mathbb{P}^3$ fibered fourfold with the two dimensional fiber coordinates $v_i$. The first four coordinates describe the $\mathbb{P}^3$ and the choice of $n_7$ and $n_9$ fix the fibration.}
\end{center}
\end{table}
From the above polyhedron one can deduce dimension one, two, three and four facets and the points within them. From that information
we can calculate the amount of Hodge numbers and in particular we find the vanishing of $h^{(2,1)}$ numbers for all $n_7$ and $n_9$.
In the Table \ref{tab:hodgeNo} we specify the Euler and Hodge numbers for the specific fibrations that allow for three-family f$G_4$-flux. We see
that only a transition between loci is possible when it is possible for the Euler number to increase.
\begin{table}[H]
\begin{center}
\begin{tabular}{|cc|c|c|c|c|c|c|} \hline
Fiber $F_i$ &Stratum $(n_7, n_9)$ & $\chi$ & $h^{(3,1)}$ & $h^{(1,1)}$ & $h^{(2,1)}$ & $h^{(2,2)}_v$ & $h^{2,2}$ \\ \hline
$F_{11}$ & $(5,6)$ & 1134 & 175 & 6 & 0 & 7 & 768 \\
$F_{11}$ & $(2,5)$ & 1554 & 245 & 6 & 0 & 7 & 1048 \\ \hline
$F_{13}$ & $(5,6)$ & 1080 & 165 & 7 & 0 & 8 & 732 \\ \hline
$F_{16}$ & $(3,6)$ & 1062 & 161 & 8 & 0 & 9 & 720 \\
$F_{16}$ & $(2,9)$ & 2274 & 363 & 8 & 0 & 9 & 820 \\ \hline
\end{tabular}
\caption{\label{tab:hodgeNo}The Euler number and Hodge numbers for all inequivalent strata that support three family fluxes.}
\end{center}
\end{table}


\bibliographystyle{utphys}	
\bibliography{ref}

\end{document}